\documentclass[aps,11pt,prd,longbibliography]{revtex4-1}

\usepackage{feynmp-auto}
\usepackage{amsmath}
\usepackage[utf8x]{inputenc} 

\usepackage{bbm}
\usepackage{soul}	
\usepackage{xcolor}

\usepackage{epsfig}
\usepackage{color}
\usepackage{slashed}
\usepackage{comment}
\usepackage{epstopdf}
\usepackage{array}
\usepackage{booktabs}
\usepackage{mathrsfs}
\usepackage[euler]{textgreek}
\usepackage{amsmath}
\usepackage{amsthm}
\usepackage{amsfonts}
\usepackage{amssymb}
\usepackage{graphicx}
\graphicspath{ {Figures/} }
\usepackage[font={small}]{caption}
\usepackage{subcaption}
\usepackage{float}
\usepackage{multirow}
\usepackage[export]{adjustbox}
\usepackage[hang, flushmargin,bottom]{footmisc} 
\usepackage{hyperref}
\hypersetup{
     colorlinks   = true,
     citecolor    = blue
}
\usepackage{placeins}
\usepackage{enumitem}
\usepackage{slashed}
\usepackage{bbm}
\usepackage{appendix}

\usepackage{titlesec}
\titleformat{\chapter}[display]
  {\normalfont\LARGE\bfseries}
  {\chaptertitlename\ \thechapter}{5pt}{\LARGE}
  \titlespacing*{\chapter}{0pt}{-20pt}{35pt}
\usepackage{bigstrut}
\usepackage{pst-node}
\usepackage{pstricks}
\usepackage{physics}
\usepackage{mathtools}
\usepackage{setspace}
\usepackage{fancyhdr}
\usepackage[makeroom]{cancel}
\usepackage{feyn}
\newcommand{\be}{\begin{equation}}
\newcommand{\ee}{\end{equation}}
\newcommand{\bes}{\begin{equation*}}
\newcommand{\ees}{\end{equation*}}

\usepackage{hyperref}
\hypersetup{%
  colorlinks = true,
  linkcolor  = black
}
\usepackage{soul}

\newcommand{\beq}{\begin{equation}}
\newcommand{\eeq}{\end{equation}}

\newcommand{\U}{\,{\rm U}}

\DeclareUnicodeCharacter{2212}{-}

\AtBeginDocument{%
    \newwrite\bibnotes
    \def\bibnotesext{Notes.bib}
    \immediate\openout\bibnotes=\jobname\bibnotesext
    \immediate\write\bibnotes{@CONTROL{REVTEX41Control}}
    \immediate\write\bibnotes{@CONTROL{%
    apsrev41Control,author="08",editor="1",pages="1",title="0",year="1"}}
     \if@filesw
     \immediate\write\@auxout{\string\citation{apsrev41Control}}%
    \fi
}%

\begin{document}

\title{\Large {\bf{Probing the Nature of Neutrinos with a New Force}}}
\author{Pavel \surname{Fileviez P\'erez}}
\email{pxf112@case.edu}
\author{Alexis D. \surname{Plascencia}}
\email{alexis.plascencia@case.edu}
\affiliation{\vspace{0.15cm} \small{Physics Department and Center for Education and Research in Cosmology and Astrophysics
(CERCA), Case Western Reserve University, Cleveland, OH 44106, USA}}

\vspace{0.5cm}

\begin{abstract}
We discuss the possibility to distinguish between Dirac and Majorana neutrinos in the context of the minimal gauge theory for neutrino masses, the $B-L$ gauge extension of the Standard Model. We revisit the possibility to observe lepton number violation at the Large Hadron Collider and point out the importance of the decays of the new gauge boson to discriminate
between the existence of Dirac or Majorana neutrinos. 
\end{abstract}
\maketitle 

\hypersetup{linkcolor=blue}
\section{INTRODUCTION}
After the discovery of the Brout-Englert-Higgs boson at the Large Hadron Collider (LHC), we understand how the electroweak symmetry is broken in nature and how all the charged fermions should acquire mass through the Higgs mechanism. Unfortunately, the Standard Model (SM) does not provide an explanation for the origin of neutrino masses, and hence, the experimental evidence of neutrino masses calls for new physics beyond the SM.  Thanks to the effort of the experimental community, the mixing angles and mass splittings in the neutrino sector have been measured with good precision, see Ref.~\cite{Tanabashi:2018oca} for a review about neutrino physics. 

The nature of neutrinos remains unknown and it is a central open question in particle physics. The neutrinos can be either Dirac or Majorana fermions~\cite{Majorana:1937vz}, in the Dirac case the anomaly-free symmetry $B-L$ is conserved or broken in a unit larger than two, while in the Majorana case $B-L$ is broken in two units. The simplest way to distinguish 
between the existence of Dirac or Majorana neutrinos is to search for exotic lepton number $({\rm{or}} \ B-L)$ violating processes at low energies such as neutrinoless double beta decays~\cite{Racah:1937qq,Furry:1939qr} or for signatures at colliders~\cite{Keung:1983uu} if the $B-L$ breaking scale is low. Clearly, the discovery of any of these processes will be crucial to understand the origin of neutrino masses and complete our understanding about mass generation.

The simplest mechanism for Majorana neutrino masses is the canonical seesaw mechanism~\cite{Minkowski:1977sc,GellMann:1980vs, Mohapatra:1979ia,Yanagida:1979as}. 
In this context, by adding at least two copies of right-handed neutrinos it is possible to generate neutrino masses in agreement with the experimental observations. Unfortunately, the seesaw 
scale could be very large, $M_{\rm{seesaw}} \leq 10^{14-15}$ GeV, and we might not have direct access to the mechanism behind neutrino masses. However, if the seesaw scale lies below or near the TeV scale, then we might test this mechanism in the near future. In the absence of 
any signature from neutrinoless double beta decays we need to investigate the possibility to observe lepton number violating signatures at particle colliders to establish the nature of neutrinos.

If the relevant scale ($B-L$ scale) for the generation of Majorana neutrino masses is relatively close to the electroweak scale we can hope to observe signatures with same-sign 
leptons at colliders~\cite{Keung:1983uu}. In the context of the canonical seesaw mechanism, there have been different studies of the production of right-handed neutrinos at particle colliders, see for example the studies in Refs.~\cite{Han:2006ip, Kersten:2007vk, delAguila:2008cj, delAguila:2008hw, Atre:2009rg, Khachatryan:2015gha, Aad:2015xaa, Khachatryan:2016olu}. The main production channel considered in many studies is $p p \to W^\pm \to N \ell^\pm$, which is generically suppressed by the active-sterile neutrino mixing. In the scenario where the right-handed neutrino masses are below the $W$-gauge boson mass, $M_N<M_W$, the right-handed neutrinos could be discovered using displaced vertices~\cite{Helo:2013esa, Izaguirre:2015pga, Batell:2016zod}. Furthermore, if $M_N<M_h/2$ the SM Higgs boson can decay into a pair of right-handed neutrinos and constraints can be placed by studying the properties of the Higgs~\cite{Graesser:2007yj, Caputo:2017pit, Deppisch:2018eth, Butterworth:2019iff}. For more details and a complete list of references see the review in Ref.~\cite{Cai:2017mow}.

The simplest gauge theory for neutrino masses corresponds to promoting $B-L$ to a local symmetry. This is because the three right-handed neutrinos automatically cancel all gauge anomalies. 
In this context the neutrinos are Dirac fermions if the new $Z_{BL}$ gauge boson acquires mass through the Stueckelberg 
mechanism~\cite{Feldman:2011ms} or if $B-L$ is spontaneously broken in more than two units. Alternatively, a new scalar with $B-L$ charge equal to two can be introduced to break $B-L$ spontaneously and generate Majorana masses for the neutrinos via the seesaw mechanism. In this scenario, the $Z_{BL}$ can mediate the pair production of right-handed neutrinos, and hence, this channel has the advantage of not being suppressed by the active-sterile neutrino mixing~\cite{Perez:2009mu}. See Refs.~\cite{Huitu:2008gf, Basso:2008iv, Kang:2015uoc} for other studies along these lines.

In this article, we discuss the possibility to distinguish between Dirac and Majorana neutrinos in the context of the minimal gauge theory for neutrino masses based on $B-L$.
We revisit the possibility to observe lepton number violation at the LHC and point out the importance of measuring the decay branching ratios of the new gauge boson to discriminate 
between the existence of Dirac or Majorana neutrinos. Clearly, a future simultaneous discovery of the $Z_{BL}$ gauge boson and heavy right-handed neutrinos will be evidence for Majorana neutrinos. However, when $M_N> M_{Z_{BL}}/2$ the production cross-section for a pair of right-handed neutrinos is highly suppressed and the prospects for observing lepton number violation are very small. We show how to distinguish between Majorana and Dirac neutrinos if a $Z_{BL}$ gauge boson is discovered even if there is no direct discovery of the right-handed neutrinos.

The structure of our work is the following: in Section~\ref{sec:U1BL}, we discuss the current bounds on the $\U(1)_{B-L}$ gauge boson mass and its coupling to matter. 
In Section~\ref{sec:LNV}, we revisit the lepton number violating signals at the LHC through the process $pp\to NN\to l^\pm l^\pm 4j$, 
we show the predictions for the latter 
by performing the most general analysis. In Section~\ref{sec:DiracMajorana}, we demonstrate how to distinguish between Dirac and Majorana neutrinos by measuring the decay width of $Z_{BL}$. We present our summary in Section~\ref{sec:Summary}.

\section{MINIMAL GAUGE THEORY FOR NEUTRINO MASSES}
\label{sec:U1BL}
The simplest gauge theory for neutrino masses is based on the local $B-L$ gauge symmetry. The right-handed neutrinos needed to generate Dirac/Majorana masses for 
neutrinos are also the extra degrees of freedom needed to define an anomaly-free gauge theory based on $B-L$. In this context the new gauge boson, $Z_{BL}$, has the following 
interactions:
\begin{equation}
{\cal{L}} \supset g_{BL}  \left( \overline{e_i} \gamma^\mu e_i - \frac{1}{3}  \overline{u_i} \gamma^\mu u_i - \frac{1}{3}  \overline{d_i} \gamma^\mu d_i 
+ \overline{\nu_i}_L \gamma^\mu \nu_{iL} + \overline{\nu_i}_R \gamma^\mu \nu_{iR} \right) Z^{BL}_\mu,
\end{equation}
where the family index $i=1,2,3$. In the above equation $e_i=e_{iL}+e_{iR}$, $u_i=u_{iL}+u_{iR}$, and $d_i=d_{iL}+d_{iR}$ are the 
Dirac spinors for the charged fermions. 

\begin{figure}[h]
\centering
\includegraphics[width=0.67\linewidth]{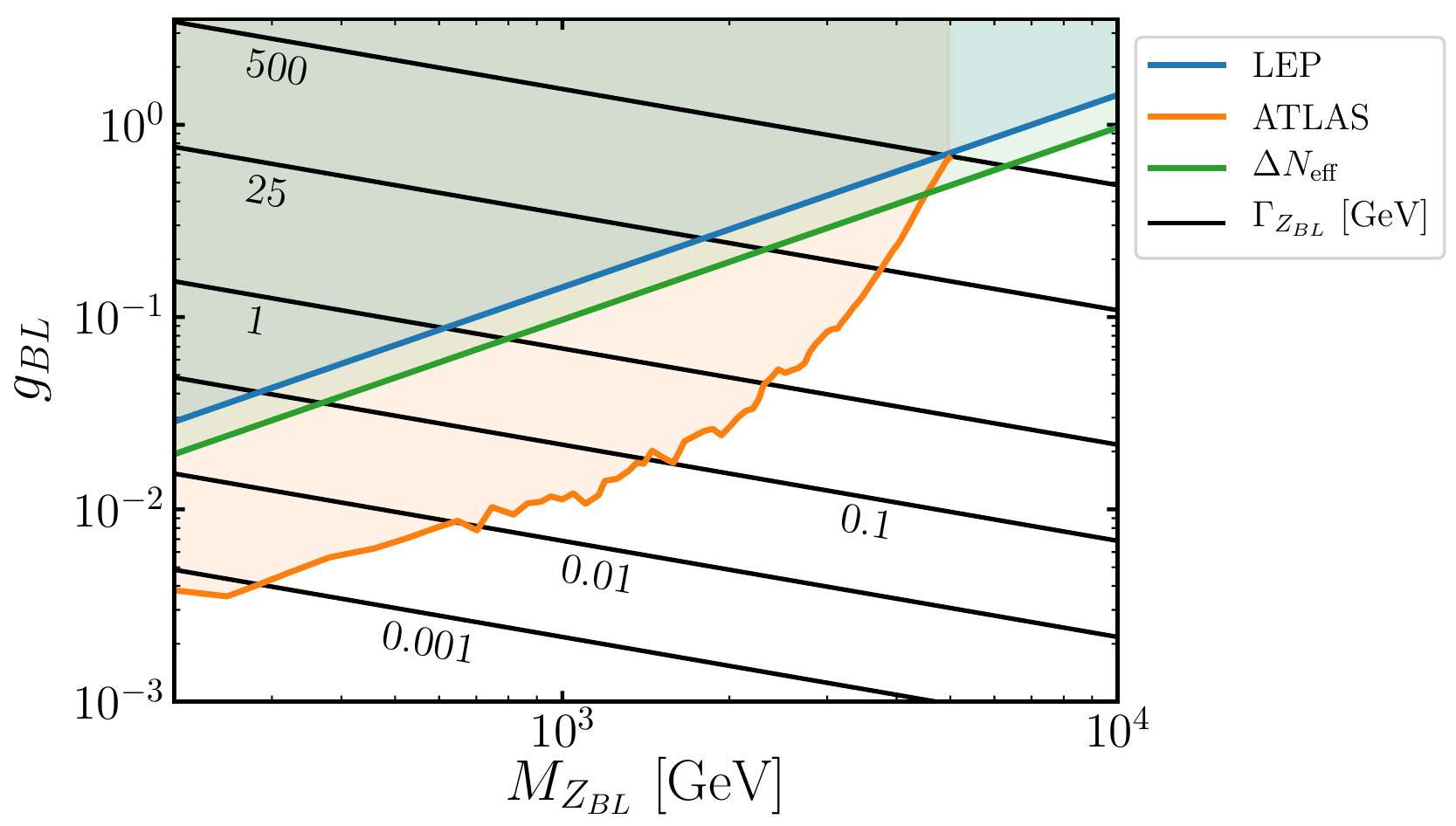}
\caption{Collider bounds in the $g_{BL}-M_{Z_{BL}}$ plane. The blue line corresponds to the bound from LEP~\cite{Alioli:2017nzr}, while the orange line corresponds to dilepton searches at the LHC with $\sqrt{s}=13$ TeV and 36.1 fb$^{-1}$ by the ATLAS collaboration~\cite{Aaboud:2017buh}. 
The constraint from the bound on $\Delta N_{\rm eff}$~\cite{FileviezPerez:2019cyn} is shown by the green line and  applies only to the scenario with Dirac neutrinos. The predictions for the decay width of $Z_{BL}$ in the scenario with Dirac neutrinos are shown by the black lines. }
\label{bounds}
\end{figure}

In Fig.~\ref{bounds} we show the relevant bounds in the $g_{BL}-M_{Z_{BL}}$ plane. The blue line corresponds to the bound from LEP~\cite{Alioli:2017nzr} ($M_{Z_{BL}}/g_{BL}>7$ TeV), 
while the orange line corresponds to dilepton searches at the LHC with $\sqrt{s}=13$ TeV and 36.1 fb$^{-1}$ \cite{Aaboud:2017buh}. The black lines define the different 
values for the decay width of the $Z_{BL}$ gauge boson, and the bound from $N_{\rm eff}$~\cite{FileviezPerez:2019cyn} ($M_{Z_{BL}}/g_{BL}>10.33$ TeV) is shown by the green line that is only relevant when the neutrinos are Dirac fermions. Notice that the LHC bounds are the most relevant when the gauge boson mass is below 4 TeV. All these bounds are relevant to understand the predictions for the processes investigated in the next section.

In the minimal $B-L$ gauge theory the Dirac Yukawa coupling for neutrinos reads as
\begin{equation}
{\cal{L}}  \supset - Y^D_\nu  \bar{\ell}_L \tilde{H} \nu_R + \rm{h.c.},
\end{equation}
with $\ell_L \sim (\mathbf{1},\mathbf{2},-1/2)$, $\tilde{H}=i \sigma_2 H^*$, and $H \sim (\mathbf{1},\mathbf{2},1/2)$ is the SM Higgs doublet.
As we mentioned above, the mass of the $B-L$ gauge boson can be generated via the Stueckelberg mechanism~\cite{Feldman:2011ms} 
leaving the $\U(1)_{B-L}$ gauge group unbroken. 
In this simple theory the neutrinos are Dirac particles, see Ref.~\cite{Perez:2017qns} for a recent discussion of the different possibilities. 
Alternatively, a scalar can be introduced with $B-L$ charge equal to two, and once this scalar acquires a non-zero vacuum expectation value 
it will give mass to the $Z_{BL}$ and the right-handed neutrinos. Therefore, the canonical seesaw mechanism for Majorana neutrinos can be implemented.

\begin{itemize}

\item{Dirac Neutrinos}

In the case when the neutrinos are Dirac fermions the decay width of the $B-L$ gauge boson can be predicted as function of the gauge coupling and its mass. 
The branching ratio for the invisible decay can be quite large due to the fact that there is an extra contribution of the right-handed neutrinos, the invisible branching ratio is close 
to $38 \%$ as it is shown in the left panel in Fig.~\ref{fig:BrsDirMaj}. This is a simple but important result because as we will discuss in the next sections, the branching ratios of $Z_{BL}$ can be used to distinguish between the Dirac and Majorana scenarios.
\begin{figure}[h]
\centering
\includegraphics[width=0.45\linewidth]{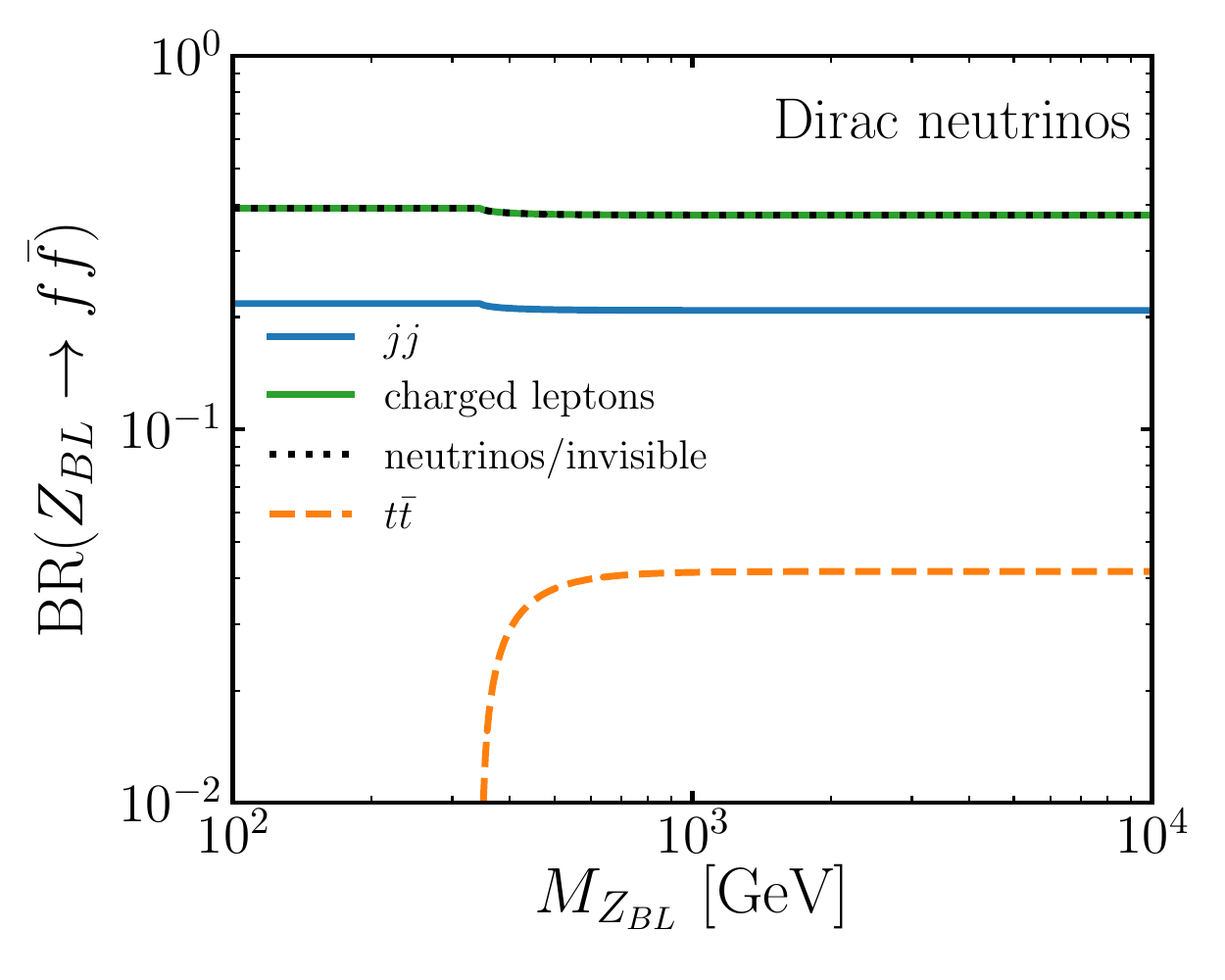}
\includegraphics[width=0.45\linewidth]{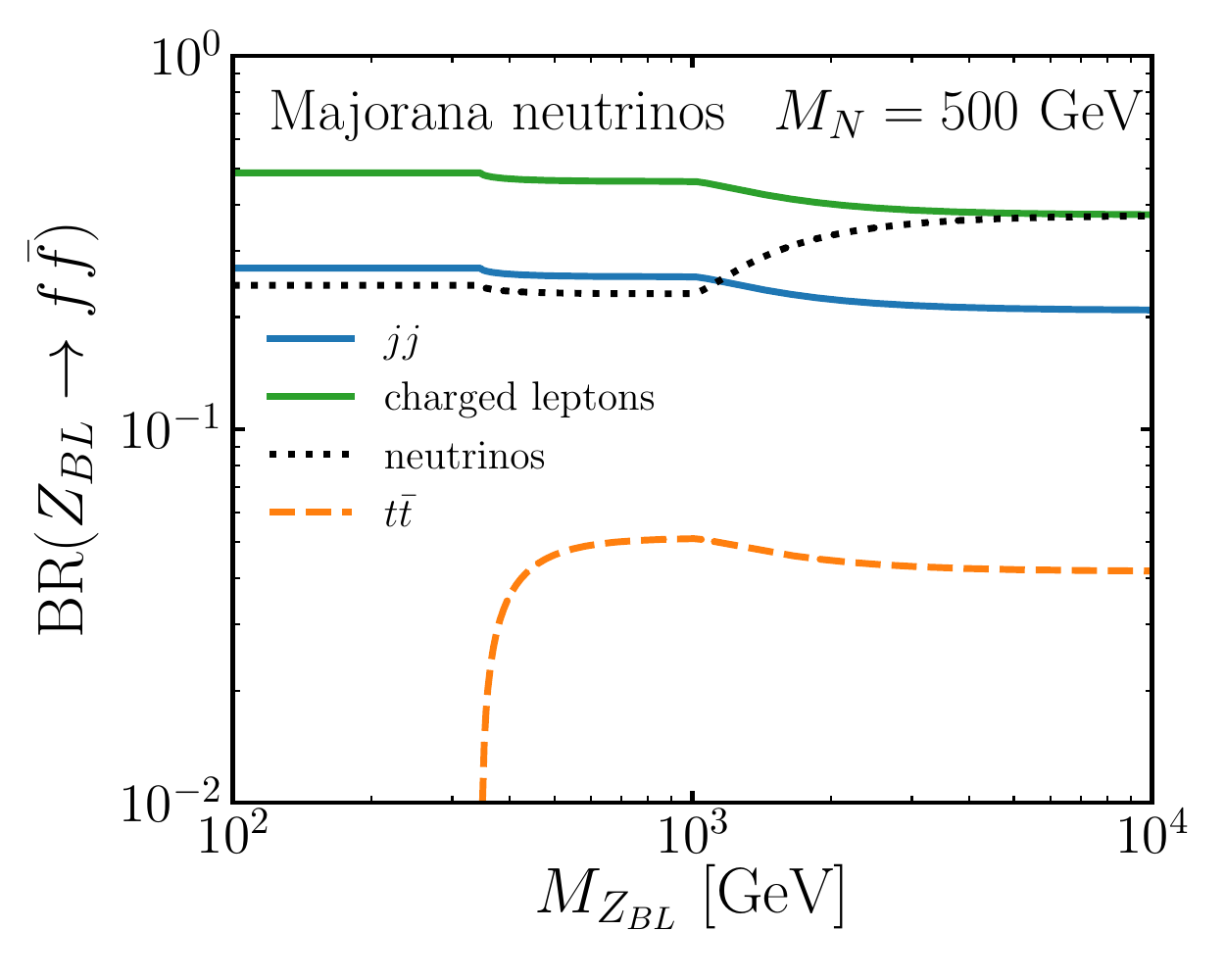}\caption{Branching ratios for the decay of $Z_{BL}$ into the different channels. The green line corresponds to the decay channels $Z_{BL} \to e^- e^+, \, \mu^- \mu^+,\, \tau^- \tau^+$, the blue line is the decay into quarks, $Z_{BL} \to q\bar{q}$, with $q=u,d,c,s,b$ and the dashed orange line is for the $Z_{BL} \to t \bar{t}$ decay. The black dotted line is the decay into Dirac neutrinos $Z_{BL} \to \nu_i \bar{\nu}_i$  (left panel) and the decay into Majorana neutrinos $Z_{BL} \to \nu_i \nu_i, \, N_i N_i$ (right panel).}
\label{fig:BrsDirMaj}
\end{figure}
\clearpage
\item{Majorana Neutrinos}

In the case with Majorana neutrinos and the canonical seesaw mechanism we can hope to observe  lepton number violation at the LHC.
The masses for Majorana neutrinos are generated after symmetry breaking through the canonical seesaw using the terms:
\begin{equation}
{\cal{L}}  \supset  -Y^D_\nu \bar{\ell}_L \tilde{H} \nu_R -  \frac{1}{2} M_N \nu_R^T C \nu_R + \rm{h.c.}
\end{equation}
Lepton number violation can be observed at the LHC through the pair production of right-handed neutrinos, i.e. $pp \to Z_{BL} \to N_i N_i$, where $N_i$ correspond to the physical states associated to the right-handed neutrinos.
In the next section, we will revisit the predictions for lepton number violation and discuss the possibility to observe these signatures.
It is important to mention that in this case the prediction for the neutrino branching ratio depends on the mass of the right-handed neutrinos.

The right panel in Fig.~\ref{fig:BrsDirMaj} shows the branching ratios of $Z_{BL}$ in the Majorana case with all three right-handed neutrino masses set to $M_N=500$ GeV. The neutrino branching ratio goes from $23\%$ to $38\%$ as the $Z_{BL} \to N_i N_i$ decay channel become kinematically allowed. Notice that the latter is not an invisible decay since the $N_i$'s can decay into visible states inside the detector.

\end{itemize}
%
\section{$B-L$ FORCE AND LEPTON NUMBER VIOLATION AT THE LHC}
\label{sec:LNV}
The observation of lepton number violation by two units at the LHC will shed light on the origin of neutrino masses. In the gauged $\U(1)_{B-L}$ scenario, the right-handed neutrinos can be produced at the LHC through the $B-L$ gauge boson: $pp \to Z_{B
L}^* \to N_i N_i$~\cite{Perez:2009mu}, with $i=1,2,3$. The cross-section for this process is given by
\begin{equation}
\sigma (pp \to N_i N_i) (s) = \int_{\tau_0}^{1} d\tau \frac{d {\cal{L}}^{pp}_{q \bar{q}}}{d \tau} \ \sigma ( q \bar{q} \to N_i N_i) (\hat{s}),
\end{equation}
where the partonic cross-section corresponds to
\begin{eqnarray}
\sigma(q \bar{q} \to Z_{BL}^* \to N_i N_i) (\hat{s})&=& \frac{g_{BL}^4}{648 \pi \hat{s}} \frac{ \left(\hat{s} - 4 M_{N_i}^2 \right)^{3/2} (2 m_q^2 + \hat{s})}{\sqrt{\hat{s} - 4 m_q^2} \left( M_{Z_{BL}}^2 \Gamma_{Z_{BL}}^2 + (\hat{s} - M_{Z_{BL}}^2)^2\right)},
 \end{eqnarray}
 and
\begin{equation}
 \frac{d {\cal{L}}^{AB}_{ab}}{d \tau} = \frac{1}{1 + \delta_{ab}} \int_\tau^1  \frac{d x}{x} \left[  f_{a/A} (x, \mu)  f_{b/B} \left(\frac{\tau}{x}, \mu\right) + f_{b/A} \left(\frac{\tau}{x}, \mu\right)  f_{a/B} (x, \mu) \right].
\end{equation} 
The parameter $\tau=\hat{s}/s$, where $\hat{s}$ is the partonic center-of-mass energy squared, $s$ is the hadronic center-of-mass energy squared, $\tau_0=4M_{N_i}^2/s$ is the production threshold, and $\mu$ is the factorization scale that is set to $\mu=M_{Z_{BL}}$. The $f$-functions correspond to the parton distribution functions for which we use the MSTW2008~\cite{Martin:2009iq} set.
\begin{figure}[t]
\centering
\includegraphics[width=0.56\linewidth]{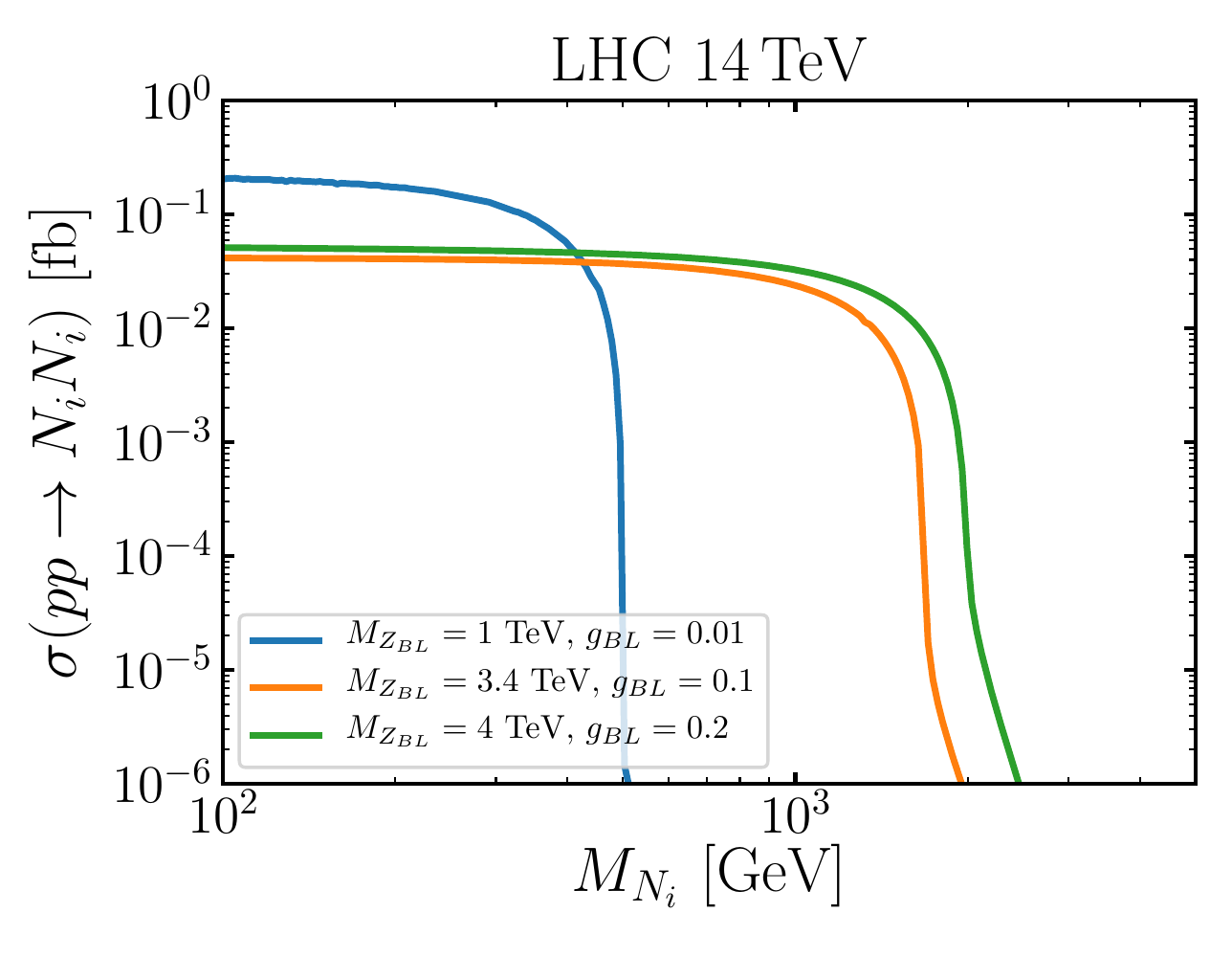}
\caption{Predictions for the production cross-section $pp \to N_i N_i$ at $\sqrt{s}=14$ TeV LHC as a function of $M_{N_i}$. Different colors correspond to different choice of parameters.}
\label{cross-section}
\end{figure} 

In Fig.~\ref{cross-section} we show the predictions for the $pp\to N_i N_i$ cross-section as a function of the right-handed neutrino mass. This plot shows that when the decay $Z_{BL} \to N_i N_i$ is kinematically closed, i.e. $M_{N_i} > M_{Z_{BL}}/2$, the cross-section drastically drops to very small values. This occurs because for these masses the cross-section never hits the $Z_{BL}$ resonance. Requiring the Majorana Yukawa coupling to be perturbative translates as an upper bound of $M_N < \sqrt{2 \pi} M_{Z_{BL}}/g_{BL}$, we make sure this is satisfied in Fig.~\ref{cross-section}.
\begin{figure}[tbp]
\centering
\includegraphics[width=0.45\linewidth]{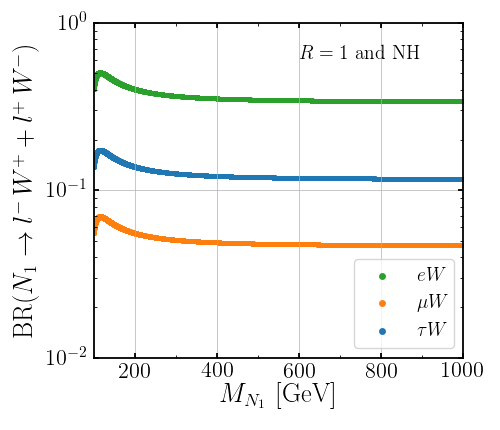}
\includegraphics[width=0.45\linewidth]{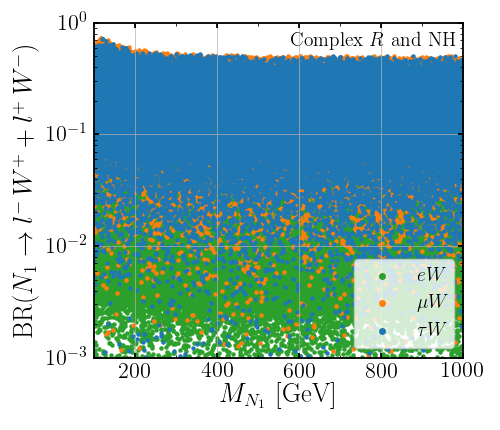}
\includegraphics[width=0.45\linewidth]{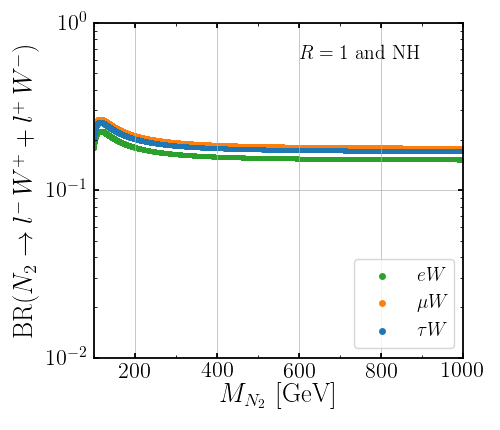}
\includegraphics[width=0.45\linewidth]{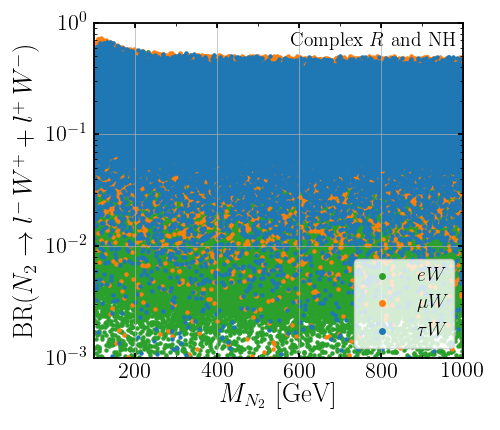}
\includegraphics[width=0.45\linewidth]{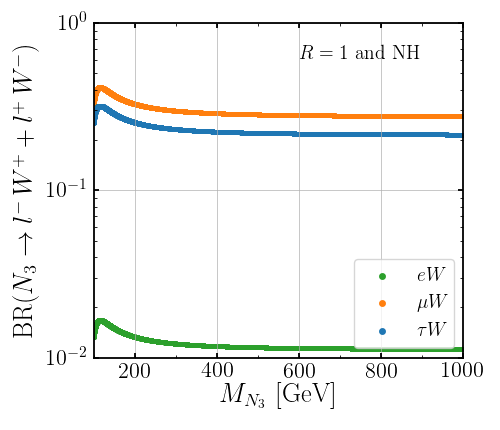}
\includegraphics[width=0.45\linewidth]{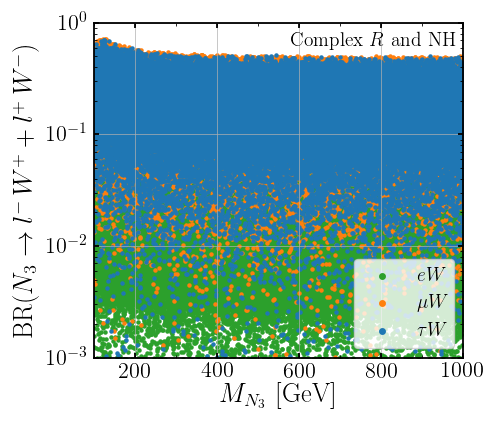}
\caption{Scatter plot of the branching ratios ${\rm BR(}N_i \to l^-\, W^+ + l^+ \, W^-)$ as a function of the right-handed neutrino mass $M_{N_i}$ and normal hierarchy. The $R$ matrix is set to the identity matrix (left panel) and a random scan is performed (right panel). The same behavior is observed for the scenario with inverted hierarchy.}
\label{fig:BRs}
\end{figure}
\begingroup
\setlength{\tabcolsep}{10pt} 
\renewcommand{\arraystretch}{1.5} 
\begin{table}[b]
\centering
\begin{tabular}{ | l | l | l | l | l | l | l |l |}
\hline
$ \sin^2 \theta_{12}$ &  $\sin^2 \theta_{23}$ &  $\sin^2 \theta_{13}$ & $\delta/^{\circ}$ & $\Delta m^2_{21}/{\rm eV}^{2}$  & $\Delta m^2_{3 1}/{\rm eV}^{2}$  (NH) & $\Delta m^2_{3 2}/{\rm eV}^{2}$ (IH) \\
 \hline 
0.310 & 0.563 & 0.02237 &  221 & $7.39 \times 10^{-5}$ & $2.528 \times 10^{-3}$ & $-2.510 \times 10^{-3} $ \\
\hline
\end{tabular}\caption{Parameters in the neutrino sector, we use the central values listed in Ref.~\cite{Esteban:2018azc}. The scenario with normal hierarchy (NH) corresponds to $\Delta m^2_{31}>0$, while the scenario with inverted hierarchy (IH) corresponds to $\Delta m^2_{32}<0$.}\label{tab:PMNS}
\end{table}
\endgroup
\begin{figure}[tbp]
\centering
\includegraphics[width=0.495\linewidth]{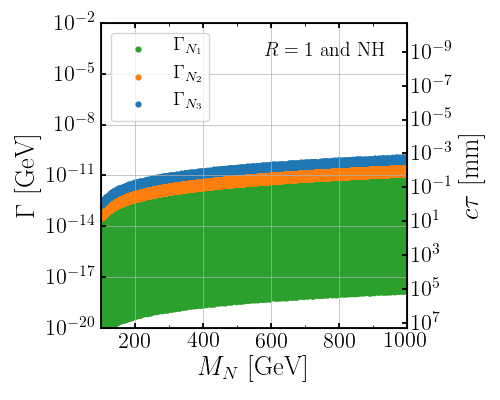}
\includegraphics[width=0.495\linewidth]{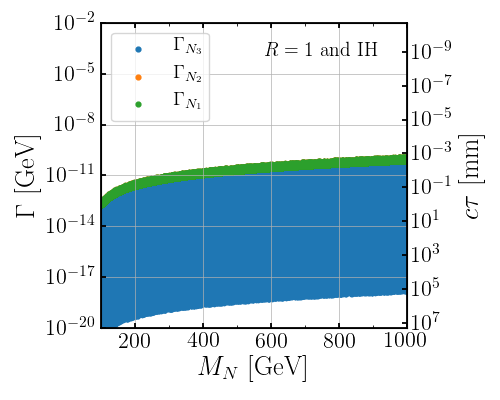}
\includegraphics[width=0.495\linewidth]{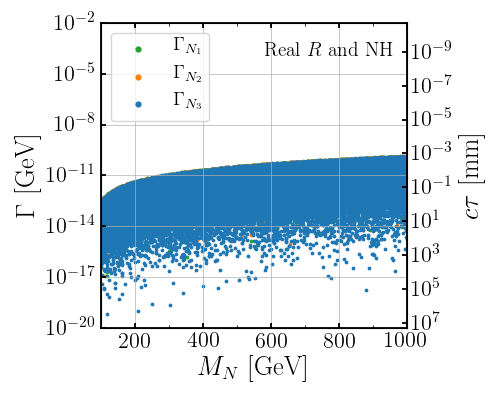}
\includegraphics[width=0.495\linewidth]{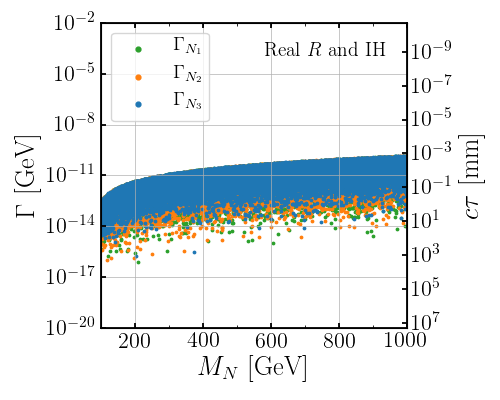}
\includegraphics[width=0.495\linewidth]{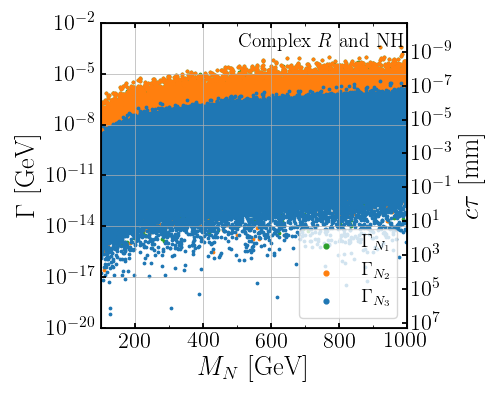}
\includegraphics[width=0.495\linewidth]{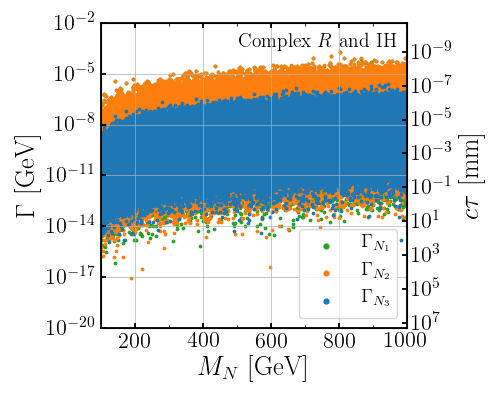}
\caption{Scatter plot of $\Gamma_N$ vs $M_N$, the green, orange and blue dots correspond to $N_1$, $N_2$ and $N_3$ respectively. For the top row we fix the $R$ matrix to the identity. For the middle (bottom) row the elements of the $R$ take on random real (complex) values. The panels on the left (right) correspond to normal (inverted) hierarchy.}
\label{fig:DecayWidth}
\end{figure}

In order to study the lepton number violating signatures we need to calculate the branching ratios for the right-handed neutrinos. The decay widths for the right-handed neutrinos are given by
\begin{align}
\Gamma(N_i \to \ell^-W^+) & = {g^2_2 \over 64\pi
M_W^2}|V_{\ell i}|^2 M_{N_i}^3 \left(1+2\frac{M_W^2}{M_{N_i}^2}\right)  \left(1-\frac{M_W^2}{M_{N_i}^2}\right)^2, \label{eq:NlW}  \\
\Gamma(N_i\to \nu_\ell Z) & = {g^2_2\over
64\pi M_W^2}|V_{\ell i}|^2 M_{N_i}^3 \left(1+2\frac{M_Z^2}{M_{N_i}^2}\right)  \left(1-\frac{M_Z^2}{M_{N_i}^2}\right)^2, \label{eq:NnuZ}  \\
\Gamma(N_i \to \nu_\ell h) & = {g^2_2\over
64\pi M_W^2}|V_{\ell i}|^2 M_{N_i}^3 \left(1-\frac{M_h^2}{M_{N_i}^2} \right)^2\cos^2\theta, \label{eq:Nnuh} 
\end{align}
where $\theta$ is the mixing angle between the SM Higgs and the scalar 
that breaks $\U(1)_{B-L}$. The matrix defining the mixing between the right-handed and left-handed neutrinos can be written 
as~\cite{Casas:2001sr}
\begin{eqnarray}
V = \ V_{\rm PMNS} \ m^{1/2} \ R \ M^{-1/2},
\label{mixing1}
\end{eqnarray}
where $V_{\rm PMNS}$ is the PMNS mixing matrix, $m={\rm{diag}} (m_1, m_2, m_3)$ is the matrix of the light neutrino masses 
and $M={\rm{diag}} (M_{N_1}, M_{N_2}, M_{N_3})$ is the matrix for the heavy neutrino masses. The $R$ matrix is complex and orthogonal, 
and it may be parametrized in terms of three complex rotation matrices
\begin{align}\label{eq:Rmatr}
R & =\begin{pmatrix}
1 & 0 & 0 \\
0 & c_{\omega_{1}} & s_{\omega_{1}} \\
0 &- s_{\omega_{1}} & c_{\omega_{1}} 
\end{pmatrix}
\begin{pmatrix}
c_{\omega_{2}} & 0 & s_{\omega_{2}} \\
0 & 1 & 0\\
-s_{\omega_{2}} & 0 & c_{\omega_{2}} 
\end{pmatrix}
\begin{pmatrix}
c_{\omega_{3}} & s_{\omega_{3}} & 0\\
-s_{\omega_{3}} & c_{\omega_{3}} & 0\\
0 & 0 & 1
\end{pmatrix},
\end{align}
where $c_{\omega_{i}}=\cos\omega_{i}$, $s_{\omega_{i}}=\sin\omega_{i}$ and $\omega_{i}$ are complex angles. The PMNS matrix can be written as
\begin{equation}
V_{\rm PMNS}= \left(
\begin{array}{lll}
 c_{12} c_{13} & s_{12} c_{13} &  s_{13} e^{-\text{i$\delta $}}
   \\
 -s_{12} c_{23} -c_{12}  s_{23} s_{13}e^{\text{i$\delta $}} \phantom{...} &
   c_{12} c_{23}-s_{12} s_{23} s_{13}  e^{\text{i$\delta $}}  &
   s_{23} c_{13}  \\
 s_{12} s_{23}-c_{12} c_{23} s_{13} e^{\text{i$\delta $}}  &
   -c_{12} s_{23} -s_{12} c_{23}  s_{13} e^{\text{i$\delta $}}\phantom{...}  &
  c_{23} c_{13} 
\end{array}
\right)\times \text{diag} (e^{i \alpha_1/2}, \, 1, \, e^{i \alpha_2/2})
\end{equation}
with $s_{ij} = \sin \theta_{ij}$, $c_{ij} = \cos \theta_{ij}$, $\delta$ is the Dirac phase and $\alpha_i$ are the Majorana phases. For their numerical values we use the central values from a recent fit~\cite{Esteban:2018azc} as given in Table~\ref{tab:PMNS}. For our numerical evaluation we perform a scan over the lightest active neutrino mass and the Majorana phases $\alpha_1$ and $\alpha_2$ in the range shown in Table~\ref{tab:ranges}. The scenario with normal hierarchy (NH) corresponds to $\Delta m^2_{3 \ell} = \Delta m^2_{31}>0$, while the scenario with inverted hierarchy (IH) corresponds to $\Delta m^2_{3 \ell} = \Delta m^2_{32}<0$.

In Fig.~\ref{fig:BRs} we show the branching ratios ${\rm BR}(N_i \to l^- W^+ + l^+ W^-)$ as a function of the right-handed neutrino mass; for the plots in the left panel the $R$ matrix is set to the identity matrix which corresponds to the simple scenario with all complex angles $\omega_i$ set equal to zero. This means that the mixings $V_{\ell i}$ depend only on low energy physics, and hence, the structure of the PMNS matrix is being reflected in these plots. Thus, by measuring the branching ratios of the $N_i$'s we can learn whether the $R$ matrix is close to the identity matrix, since in this case there is a clean prediction for each branching ratio. We focus on the $N_i \to l^\mp W^\pm $ channels because these are the ones that lead to signatures of lepton number violation as we will see below.

For the plots in the right panel in Fig.~\ref{fig:BRs} we perform a scan on the $\omega_i$ complex angles in the ranges shown in Table~\ref{tab:ranges}. The imaginary parts of $\omega_i$ exponentially enhance the entries in the $R$ matrix so we make sure that each entry in the Dirac Yukawa matrix $Y_\nu^D$ remains perturbative.
This demonstrates that once the freedom in the $R$ matrix is taken into account the predictions can change drastically. For example, the branching ratio for $N_1  \to \mu^\mp W^\pm$ which is around $4.5\%$ for $R=1$ can become as large as $50\%$ once the random scan is performed. We find that the branching ratios are not sensitive to whether we have normal hierarchy or inverted hierarchy in the active neutrino sector, so the plots have the same behavior for IH.

In Fig.~\ref{fig:DecayWidth} we show the decay width of $N_i$ as a function $M_{N_i}$ for each right handed neutrino. As can be seen from Eqs.~\eqref{eq:NlW}-\eqref{eq:Nnuh} the decay widths are proportional to the light neutrino masses, and hence, they depend on whether we have NH or IH. The plots on the left correspond to normal hierarchy while the ones on the right correspond to inverted hierarchy.  For the top row we fix $R=1$, which means the $\Gamma_{N_i}$ are only dependent on the light neutrino masses $m_{i}$. In the NH scenario this is the reason why $\Gamma_{N_1}$ can vary over several orders of magnitude, while $\Gamma_{N_2}$ and $\Gamma_{N_3}$ are restricted to a small window. For the IH scenario, the decay width $\Gamma_{N_3}$ is the one that has a large range since $\nu_3$ is the lightest.

In the middle row of Fig.~\ref{fig:DecayWidth} we show the decay width for a scan of the $R$ matrix taking only real parameters. The difference with  $R$ identity is that now there is more freedom in the decay width for $N_2$ and $N_3$ in the NH and $N_1$ and $N_2$ in the IH. Once the $R$ matrix allowed to take on random values it becomes very difficult to distinguish between the NH and the IH scenarios. An exploration for different values of the $R$ matrix is not commonly done in the literature.

In the third row of Fig.~\ref{fig:DecayWidth}  we present the results for a random scan of the $R$ matrix considering complex entries. We find that the decay width can be 6 orders of magnitude larger than when taking only real entries in the $R$ matrix; this happens because imaginary parts of $\omega_i$ exponentially enhance the entries in the mixing matrix $V$.
We find that the maximal values for the decay widths are $\Gamma^{\rm max}_{N_1} \simeq 10^{-4}$ GeV, $\Gamma^{\rm max}_{N_2} \simeq 10^{-4}$ GeV and $\Gamma^{\rm max}_{N_3} \simeq 2 \times 10^{-6}$ GeV.  Therefore, the heavy neutrinos can decay more promptly than we might expect from the naive seesaw relation $V^2 \approx m/M_N$.

The decaying length of the heavy neutrinos can have a large range from $10^{-9}$ mm to  $10^5$ mm, and hence, these heavy neutrinos can be searched for using different techniques. When the decay length is between $10^{-2}$ mm and $10^{3}$ mm then these appear at the LHC as displaced vertices. Additionally, when the lightest neutrino mass is taken to be very small, the decay length can be in the order of meters and detectors such as FASER~\cite{Feng:2017uoz} or MATHUSLA~\cite{Curtin:2018mvb} can be used to search for them.
\begingroup
\setlength{\tabcolsep}{10pt} 
\renewcommand{\arraystretch}{1.5} 
\begin{table}[tbp]
\centering
\begin{tabular}{ | l | l | }
\hline
 Parameter  &  Scan Range NH (IH) \\
 \hline 
$m_{1(3)}$ & $[10^{-9}$, $0.1]$ eV \\
$\alpha_1$	     & $\left[ -\pi, \pi  \right]$ \\
$\alpha_2$	     & $\left[ -\pi, \pi  \right]$ \\
Re$[\omega_i]$	     & $\left[ -\pi, \pi  \right]$ \\
Im$[\omega_i]$          & $\left[ -\pi, \pi  \right]$ \\
\hline
\end{tabular}\caption{Ranges of the nine free parameters in our numerical scan. For Normal Hierarchy (Inverted Hierarchy) we scan over the lightest neutrino mass $m_1$ ($m_3$).}\label{tab:ranges}
\end{table}
\endgroup

Lepton number violation can be probed by searching for the process $pp \to N_i N_i \to l^\pm l^\pm 4j$ at the LHC. The expected number of events for this process is given by
\beq
N_{\rm events} = \mathcal{L} \times \sigma(p p \to N_i N_i) \times 2 \times  {\rm BR}^2(N_i \to l^\pm W^\mp) \times {\rm BR}^2(W^\mp\to jj),
\eeq
where the hadronic decay of the $W$ boson is BR$(W^\mp\to jj)\simeq 2/3$. In Fig.~\ref{fig:Nevents} we show the expected number of events at the LHC for center-of-mass energy of 14 TeV assuming $\mathcal{L}=3000{\rm \, fb}^{-1}$ for the integrated luminosity. The points in black correspond to the simplified scenario with $R=1$ and for the gray points we perform a scan on the free parameters in the range shown in Table~\ref{tab:ranges}. See Ref.~{\cite{Perez:2009mu} where the authors discussed in detail the relevant SM backgrounds, $t\bar{t}W$ and multi-bosons, and how to distinguish between the signal and background imposing different kinematical cuts. 

The top row in Fig.~\ref{fig:Nevents} corresponds to the production and decay of $N_1$. The left panel corresponds to the $N_1 N_1 \to e^\pm e^\pm 4j$ channel. Here, the case $R=1$ is close to the largest number of events obtained from the random scan. The middle panel is for the $N_1 N_1 \to e^\pm \mu^\pm 4j$ channel and in this case the random scan can increase the number of events by a factor of 4. The right panel is for the $N_1 N_1 \to \mu^\pm \mu^\pm 4j$ channel and here the $R=1$ case predicts a much lower number of events than can be obtained from the random scan which can increase the number of events by a factor of 100. As can be appreciated, the predictions for the number of events is very sensitive to the form of the $R$ matrix and can be quite different from the ones obtained using the naive seesaw relation $V^2 \approx m_\nu/M_N$. If these channels are discovered in the near future, this information can be used to learn about the $R$ matrix and the seesaw relation. In Appendix~\ref{sec:appendix} we present the results of our scan for the active-sterile neutrino mixing. 

\begin{figure}[tbp]
\includegraphics[width=0.325\linewidth]{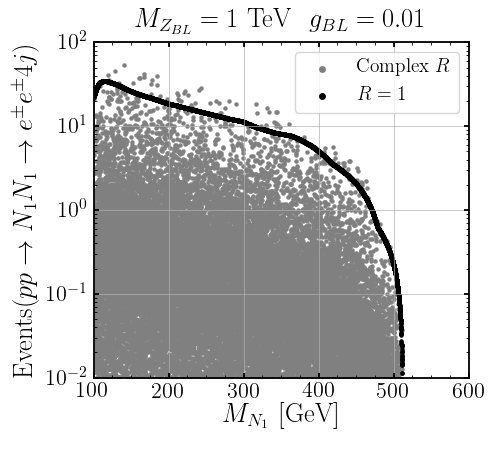}
\includegraphics[width=0.325\linewidth]{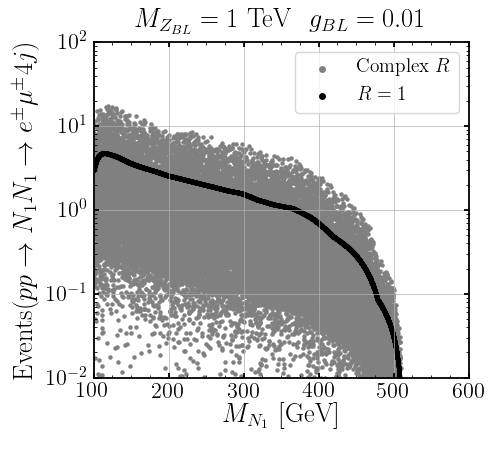}
\includegraphics[width=0.325\linewidth]{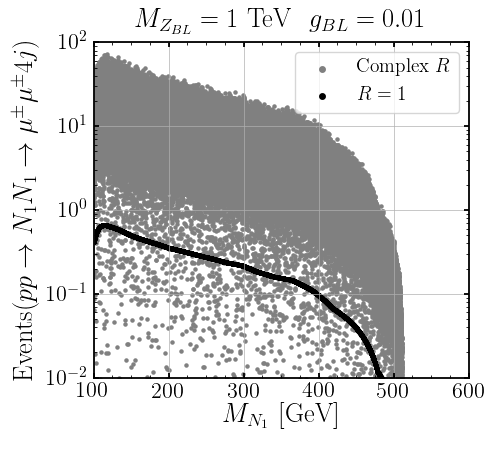}
\includegraphics[width=0.325\linewidth]{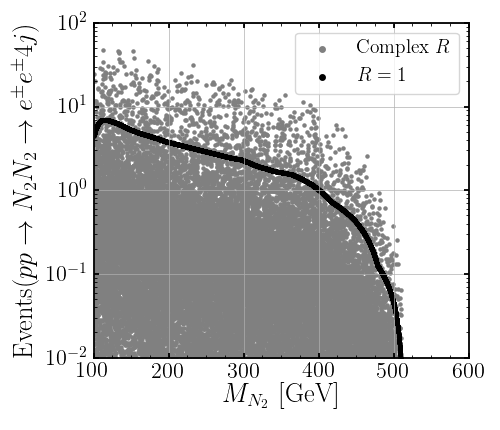}
\includegraphics[width=0.325\linewidth]{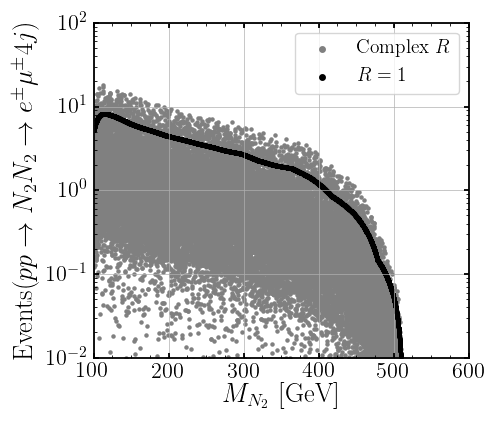}
\includegraphics[width=0.325\linewidth]{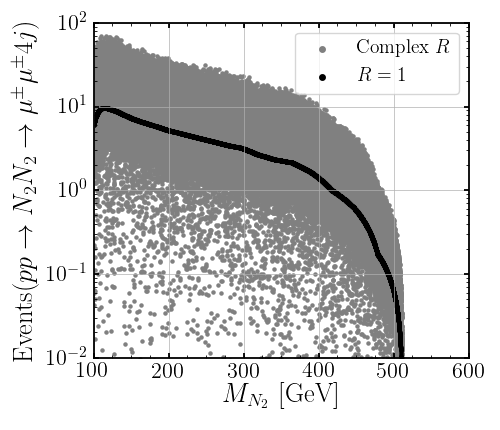}
\includegraphics[width=0.325\linewidth]{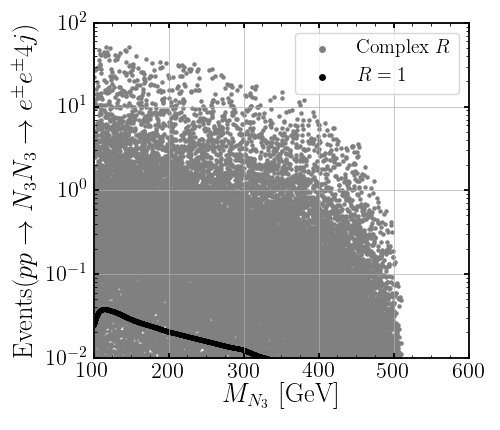}
\includegraphics[width=0.325\linewidth]{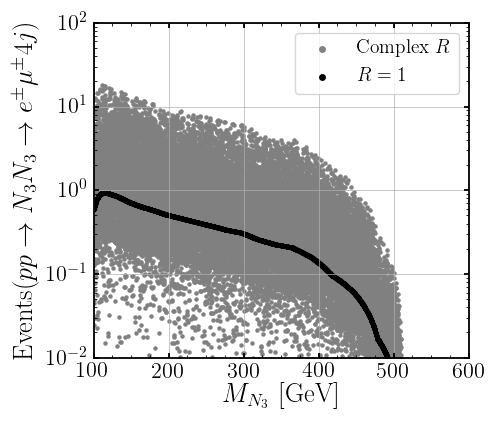}
\includegraphics[width=0.325\linewidth]{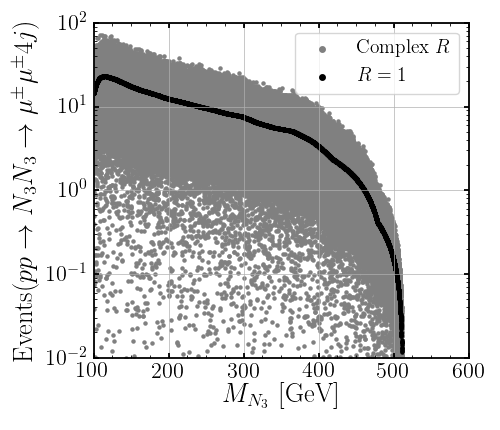}
\caption{Scatter plot of the expected number of events at the LHC for center-of-mass energy of 14 TeV assuming $\mathcal{L}=3000{\rm \, fb}^{-1}$ for the integrated luminosity. The black points correspond to the case with $R=1$, while the gray points correspond to a random scan on the entries of the $R$ matrix. These plots correspond to the case with normal hierarchy and we scan over the lightest neutrino mass, the same pattern is observed for inverted hierarchy.}
\label{fig:Nevents}
\end{figure}

\section{DIRAC vs MAJORANA: THE ROLE OF THE $Z_{BL}$ DECAY WIDTH} 
\label{sec:DiracMajorana}
The discovery of the $Z_{BL}$ gauge boson does not guarantee the discovery of right-handed neutrinos, and hence, we might be unable to disentangle between neutrinos being Dirac or Majorana. In this section, we argue that a measurement of the $Z_{BL}$ total width, $\Gamma_{Z_{BL}}$, and its decay branching ratios will suffice to distinguish between the scenario with Dirac or Majorana neutrinos. We expect the LHC to reach this precision \cite{Li:2009xh}. For example, take the high precision LEP measurement of the $Z$ boson in the SM to less than one percent $\Gamma_Z = 2.4952 \, \pm \, 0.0023$ GeV \cite{ALEPH:2005ab}.
\begin{figure}[h]
\centering
\includegraphics[width=0.55\linewidth]{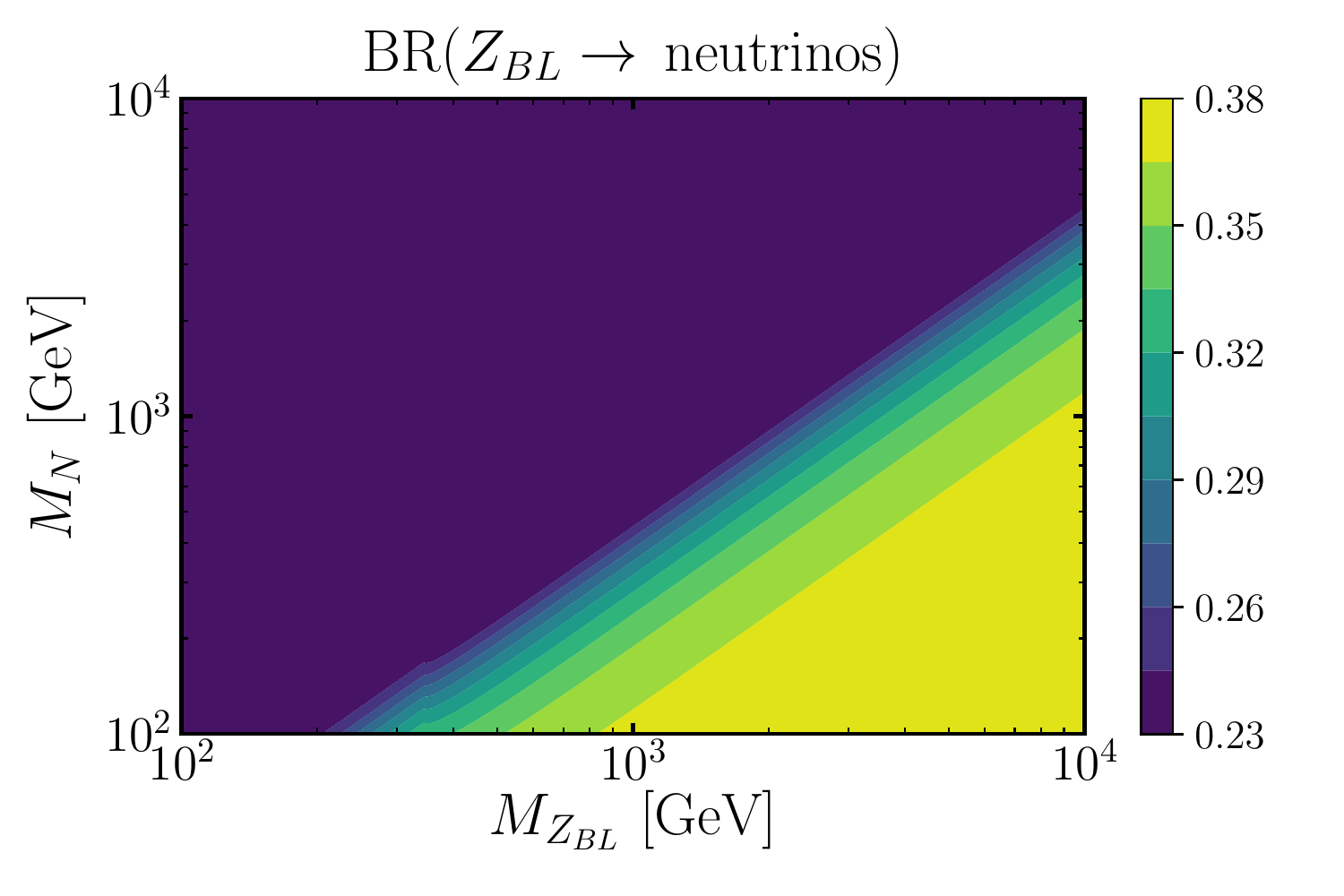}
\caption{Contour plot for the neutrino branching ratio of $Z_{BL}$ in the $M_N$ vs $M_{Z_{BL}}$ plane. This result is independent of the value of the coupling $g_{BL}$.
We take all three right handed neutrino masses $M_{N_i}$ to be equal to $M_{N}$ for simplicity.}
\label{fig:BRinv}
\end{figure}
The decay width of $Z_{BL}$  for the different channels is given by,
\beq
\Gamma_{Z_{BL}} = \Gamma_{\rm hadrons} + \Gamma_{\rm leptons} + \Gamma_\nu,
\eeq
where the last term is the contribution from the decay into neutrinos. In the scenario with Dirac neutrinos we have
\beq
\label{eq:GammaDirac}
\Gamma_\nu^D = \sum_{i=1}^{3}  \Gamma \left(Z_{BL} \to \nu_i \bar{\nu}_i \right) = 6 g_{BL}^2 \frac{M_{Z_{BL}}}{24 \pi},
\eeq
while in a scenario with Majorana neutrinos we have
\begin{align}
\Gamma_\nu^M & = \sum_{i=1}^3  \Gamma \left(Z_{BL} \to \nu_i \nu_i \right) + \sum_{i=1}^3 \Gamma \left(Z_{BL} \to  N_i N_i \right)\nonumber\\
& = 3 g_{BL}^2 \frac{M_{Z_{BL}}}{24 \pi} + \sum_{i=1,2,3} g_{BL}^2 \frac{M_{Z_{BL}}}{24 \pi} \left( 1 - \frac{4M_{N_i}^2}{M_{Z_{BL}}^2}  \right)^{3/2}. \label{eq:GammaMajorana}
\end{align}
Notice that $\Gamma_\nu^D \geq \Gamma_\nu^M$. Consequently, the total $Z_{BL}$ decay width is different depending on whether neutrinos are Dirac or Majorana.
However, we should point out that there is degeneracy $\Gamma_{\nu}^D\simeq\Gamma_{\nu}^M$ in the scenarios with $M_{N_i} \ll M_{Z_{BL}}$. 
All the results in this section are independent of the active-sterile neutrino mixing. 

In Fig.~\ref{fig:BRinv} we present a contour plot of the neutrino branching ratio for $Z_{BL}$ in the $M_N$ vs $M_{Z_{BL}}$ plane, where we are assuming that $M_{N_1}=M_{N_2}=M_{N_3}=M_{N}$ for simplicity. This branching ratio is independent of the value of the coupling $g_{BL}$. Whenever $M_N>M_{Z_{BL}}/2$ the only decays into neutrinos are $Z_{BL}\to \nu_i \nu_i$ and this branching ratio is equal to $23\%$. As the $Z_{BL}\to N_i N_i$ channels become kinematically open the neutrino branching ratio starts to increase and goes to $38\%$ for $M_N\ll M_{Z_{BL}}$. 
To quantify the difference between the neutrino width in the Dirac vs Majorana case we define the following quantity
\begin{equation}
\delta \Gamma_\nu \equiv \frac{\Gamma^D_\nu - \Gamma^M_\nu} {\Gamma^M_\nu},
\end{equation}
where $\Gamma^D_\nu$ corresponds to Dirac neutrinos given by Eq.~\eqref{eq:GammaDirac} and in the limit of massless neutrinos depends only on $M_{Z_{BL}}$, while $\Gamma^M_\nu$ which given in Eq.~\eqref{eq:GammaMajorana} corresponds to the Majorana case and depends on both $M_{Z_{BL}}$ and $M_N$. The $\delta \Gamma_\nu$ parameter can range between 0 and 1. For $M_N \ll M_{Z_{BL}}$ $\delta \Gamma_\nu$ is close to 0 and it is hard to disentangle between Dirac and Majorana. As $M_N$ approaches $M_{Z_{BL}}$ then this quantity increases and once the $Z_{BL}\to N_i N_i$ channels are closed then $\Gamma^D_\nu=2\Gamma^M
_\nu$ and we have $\delta \Gamma_\nu =1$. This behavior is manifested in Fig.~\ref{fig:deltaGammaInv}, where we show the parameter $\delta \Gamma_\nu$ as a function of the gauge boson mass for different values of $M_N$. 

\begin{figure}[h]
\centering
\includegraphics[width=0.6\linewidth]{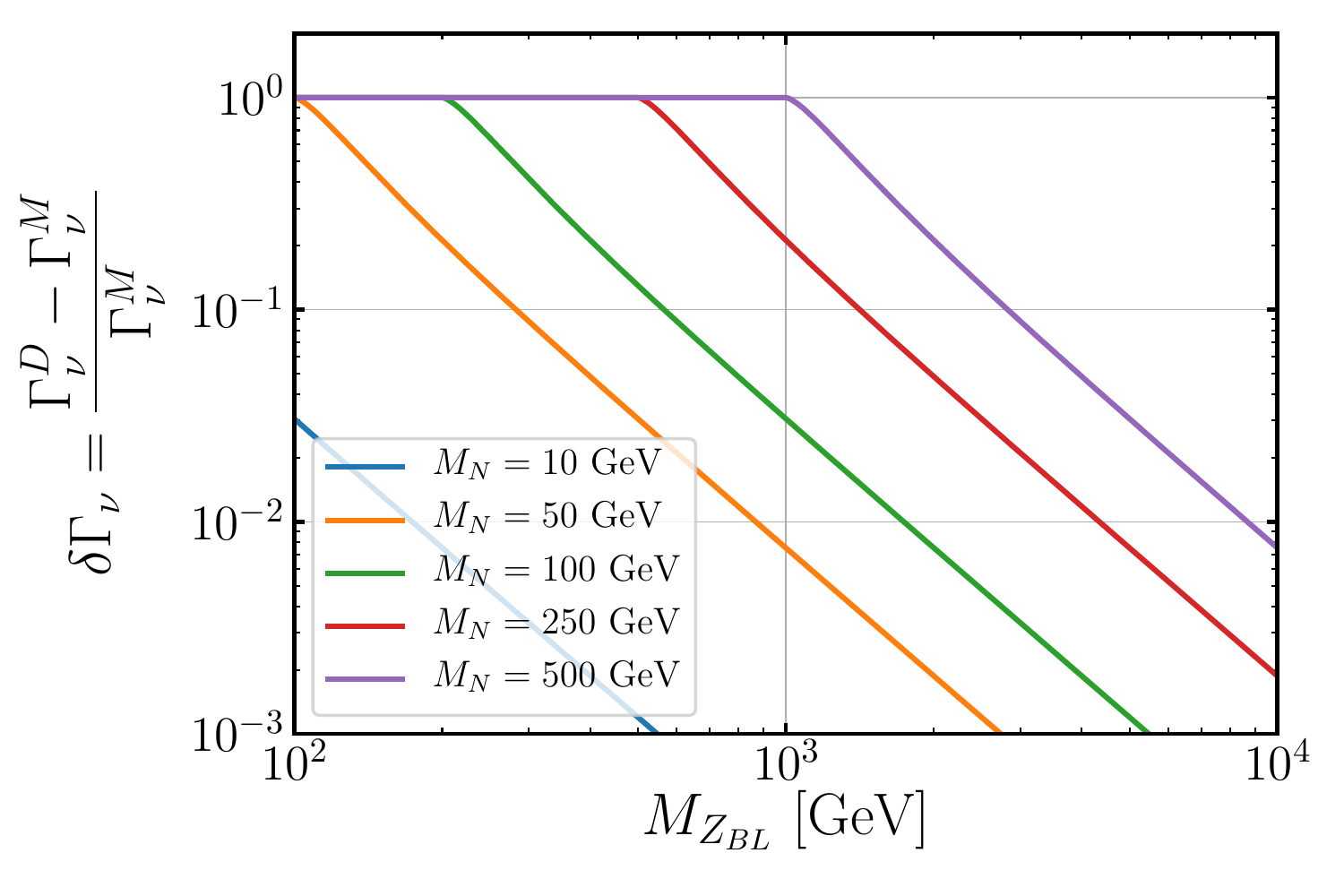}
\caption{The difference in the total $Z_{BL}$ width in the Dirac vs Majorana scenarios normalized with respect to the Majorana case as a function of the $Z_{BL}$ mass. 
These results are independent of the value of the gauge coupling $g_{BL}$.}
\label{fig:deltaGammaInv}
\end{figure}

\begin{figure}[h]
\centering
\includegraphics[width=0.6\linewidth]{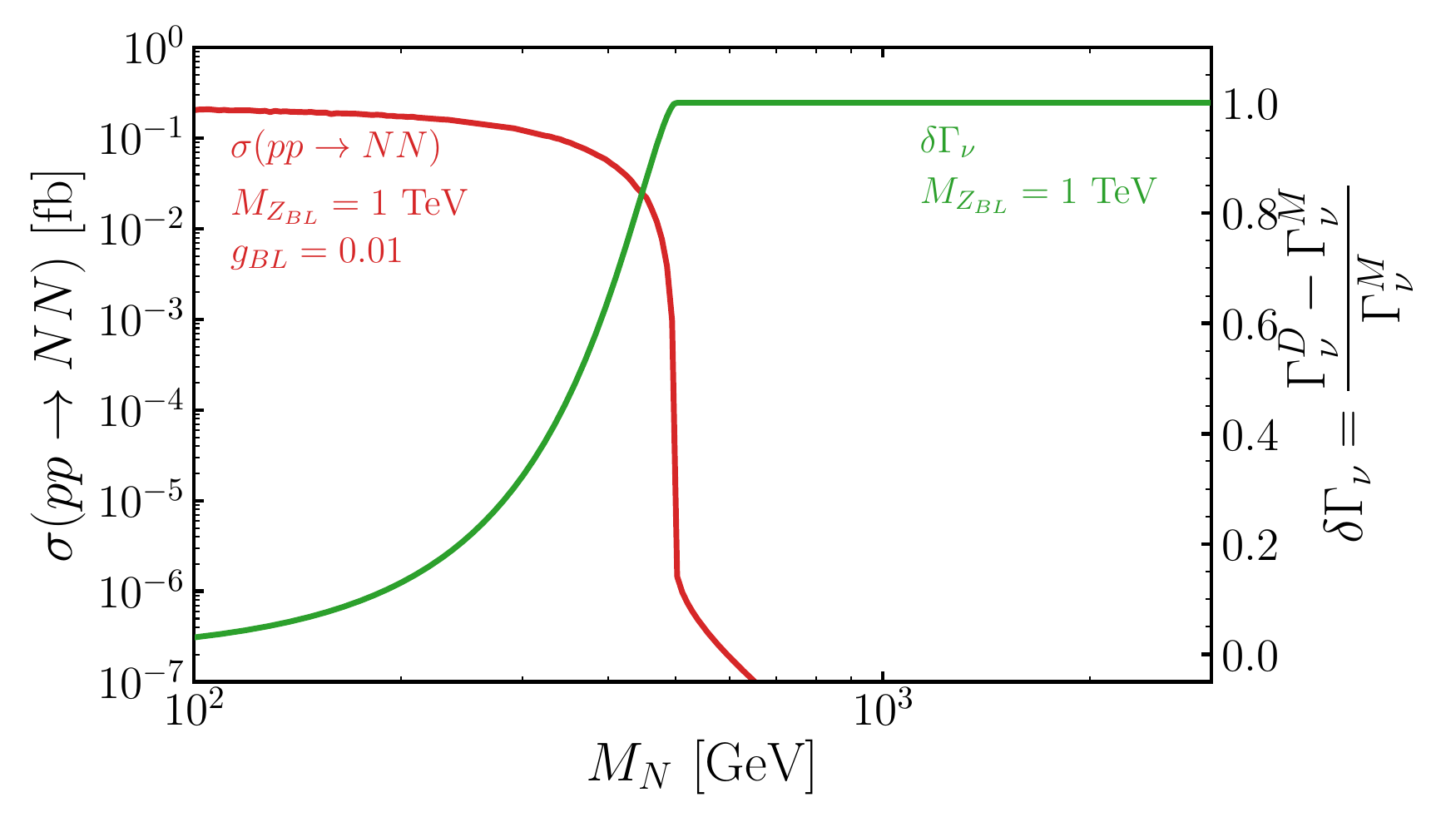}
\includegraphics[width=0.6\linewidth]{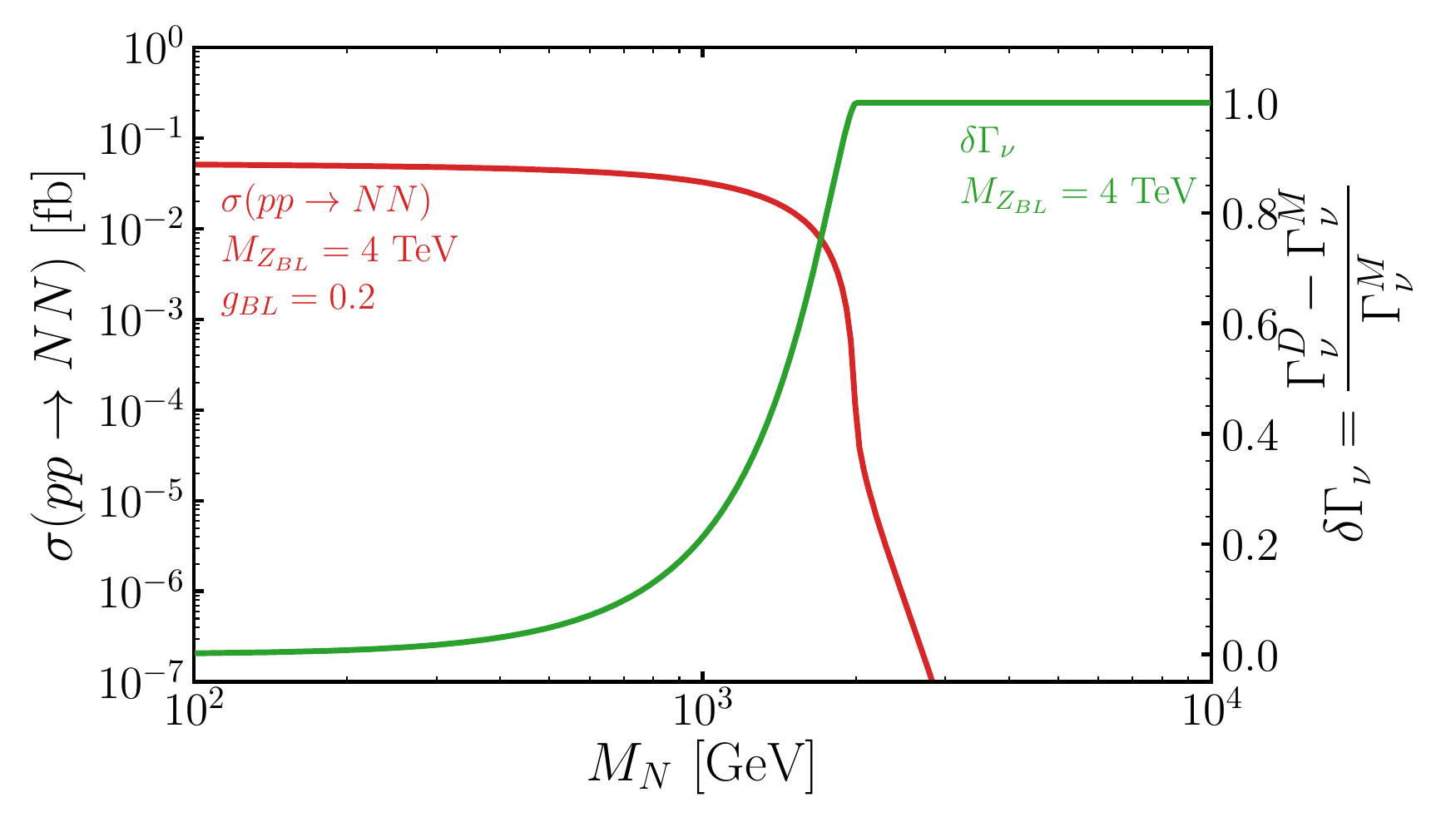}
\caption{The red line corresponds to the production cross-section of $pp\to NN$ at the LHC with center-of-mass energy of 14 TeV. The blue line is the difference in the $Z_{BL}$ neutrino width between the Dirac and the Majorana cases normalized with respect to Majorana case. Both are plotted as a function of the $M_N$. The upper (lower) panel corresponds to $M_{Z_{BL}}=1$ TeV and $g_{BL}=0.01$ ($M_{Z_{BL}}=4$ TeV and $g_{BL}=0.2$).}
\label{fig:deltaMN}
\end{figure}

Now, let us discuss the correlation between the lepton number violating processes and the decay width of the $B-L$ gauge boson. 
In Fig.~\ref{fig:deltaMN} we show in red the production cross-section for a pair of right-handed neutrinos and in green the $\delta \Gamma_{\nu}$ parameter as a function of $M_N$. These plots show the complementarity between direct production of $N_i$'s and the measurement of the $Z_{BL}$ neutrino branching ratio to distinguish between Dirac and Majorana neutrinos. These plots show that as the pair production cross-section goes down, the $\delta \Gamma_{\nu}$ parameter increases eventually becoming equal to one. 

If a $Z_{BL}$ gauge boson is discovered and the right-handed neutrinos lie in the range $100 \,\, {\rm GeV} < M_N < M_{Z_{BL}}/2$ then there is a good possibility to directly produce the  right-handed neutrinos at the LHC. However, when $M_N>M_{Z_{BL}}/2$ it becomes very difficult to produce the right-handed neutrinos; thus, we could either be in a scenario with Dirac or Majorana neutrinos. Nonetheless, by measuring the neutrino branching ratio of ${Z_{BL}}$ we can discriminate between these two possibilities. There are three different scenarios that are possible:

\begin{itemize}

\item ${\rm BR}(Z_{BL} \to {\rm neutrinos}) \simeq 23\%$: Measuring the neutrino branching ratio of $Z_{BL}$ close to $23\%$ implies that neutrinos are Majorana and that the channels $Z_{BL}\to N_i N_i$ are kinematically closed. This corresponds to having $\delta \Gamma_{\nu} = 1$. Consequently, even if we are unable to directly produce $N_i N_i$ we will have indirect evidence that neutrinos are Majorana fermions.

\item $23\% <  {\rm BR}(Z_{BL} \to {\rm neutrinos}) < 38\% $: A measurement of the neutrino branching ratio between $23\%$ and $38\%$ will mean that neutrinos are Majorana and that both channels $Z_{BL} \to \nu_i \nu_i $ and $Z_{BL}\to N_i N_i$  are open. The Majorana nature can be further confirmed by direct observation of the right-handed neutrinos at particle colliders. This corresponds to having $0 < \delta \Gamma_{\nu} < 1$.

\item ${\rm BR}(Z_{BL} \to {\rm neutrinos}) \simeq 38\% $: If the neutrino branching ratio is measured very close to $38\%$ then it becomes hard to disentangle the nature of neutrinos since we can either be in the case with Dirac neutrinos or the one with Majorana neutrinos and $M_N\ll M_{Z_{BL}}$. This corresponds to having $\delta \Gamma_{\nu} \simeq 0$.

\end{itemize}

\begingroup
\setlength{\tabcolsep}{10pt} 
\renewcommand{\arraystretch}{1.5} 
\begin{table}[h]
\centering
\begin{tabular}{ | l | l | l | l | l | }
\hline
& Dirac neutrinos &  $M_N = 10$ GeV &  $M_N=500$ GeV & $M_N=2$ TeV  \\
 \hline 
$\Gamma(Z_{BL} \to$ neutrinos)&  2.70563 GeV &  2.70556 GeV &  2.53395 GeV &  1.35282 GeV \\
 \hline
$\Gamma(Z_{BL} \to$ all)  &  7.21501 GeV & 7.21494 GeV  & 7.04333 GeV &  5.86219 GeV \\
 \hline  
\end{tabular}\caption{Numerical values in units of GeV for the decay width for the channel $Z_{BL} \to {\rm neutrinos}$ and the $Z_{BL}$ total width for four different scenarios. We set $M_{Z_{BL}}=3.4$ TeV and $g_{BL}=0.1$.}\label{tab:Scenario}
\end{table}
\endgroup

In the Dirac scenario the branching ratio into neutrinos is invisible. In the Majorana case, the decay into light neutrinos $Z_{BL}\to \nu_i \nu_i$ is always invisible. However, for the decays $Z_{BL}\to  N_i N_i$, the heavy neutrinos $N_i$ can subsequently decay into visible particles inside the detector. In Table~\ref{tab:Scenario}} we show the predictions for the $Z_{BL}$ decay into neutrinos and its total width for $M_{Z_{BL}}=3.4$ TeV and $g_{BL}=0.1$. As this table shows, when the right-handed neutrino mass is below $10$ GeV it is difficult to distinguish between Dirac and Majorana neutrinos because the difference in the decay width into neutrinos is smaller than $10^{-4}$ GeV. However, above $10$ GeV one can distinguish the two scenarios for neutrino masses.

%

\section{SUMMARY} 
\label{sec:Summary}
We have discussed how to distinguish between Dirac and Majorana neutrinos in the simplest gauge theory for neutrino masses; namely, the $B-L$ gauge extension of the SM. Assuming that the $B-L$ symmetry breaking scale is not far from the electroweak scale, we revisited 
the prospects for observing lepton number violation at the Large Hadron Collider. We performed a general random scan on the parameters in the $R$ matrix that enters in the mixing between neutrinos and demonstrated that the lifetime of the right-handed neutrinos can span many order of magnitudes. Even for right-handed neutrino masses above the electroweak scale decaying lengths in the order of meters are possible. We have shown that a large number of events for the processes $pp \to N_i N_i \to l^\pm l^\pm 4j$ can be observed in some cases 
and using these channels one can learn about the structure of the mixing matrix and the seesaw relation. 

We have discussed how the measurement of the $Z_{BL}$ decay width and branching ratios can help to discriminate between Majorana and Dirac neutrinos. Three different scenarios are possible: 
\textbf{i)} ${\rm BR}(Z_{BL} \to {\rm neutrinos})\simeq23\%$ would mean that $M_N>M_{Z_{BL}}/2$ in which the pair-production cross-section for right-handed neutrinos is highly suppressed; nonetheless, measuring this branching ratio will imply that neutrinos are Majorana. 
\textbf{ii)} $23 \% \lesssim {\rm BR}(Z_{BL} \to {\rm neutrinos})\lesssim 38\%$ means that the decay channels $Z_{BL}\to N_iN_i$ are open and we will be able to directly pair produce the right-handed neutrinos at colliders. \textbf{iii)} ${\rm BR}(Z_{BL} \to {\rm neutrinos})\simeq 38\%$ would be a pessimistic scenario in which we have $M_N \ll M_{Z_{BL}}$, which makes the right-handed neutrinos hard to observe at particle colliders and also hard to disentangle between Dirac and Majorana neutrinos, since the prediction for Dirac neutrinos is ${\rm BR}(Z_{BL} \to {\rm neutrinos})=38\%$.
Our results could help uncover whether the neutrinos are Dirac or Majorana fermions and complete our understanding of the mass generation.

\vspace{1.0cm}

{\small{\textit{Acknowledgments}: The work of P.F.P. has been supported by the U.S. Department of Energy, Office of Science, Office of High Energy Physics, under Award Number DE-SC0020443. We thank C. Murgui for discussions.}}

\clearpage
\appendix
\section{Neutrino mixings}
\label{sec:appendix}
\begin{figure}[h]
\includegraphics[width=0.325\linewidth]{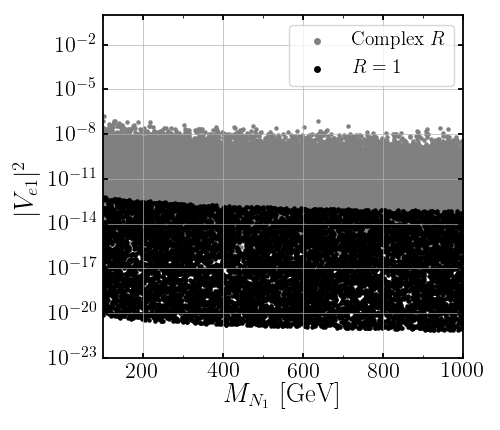}
\includegraphics[width=0.325\linewidth]{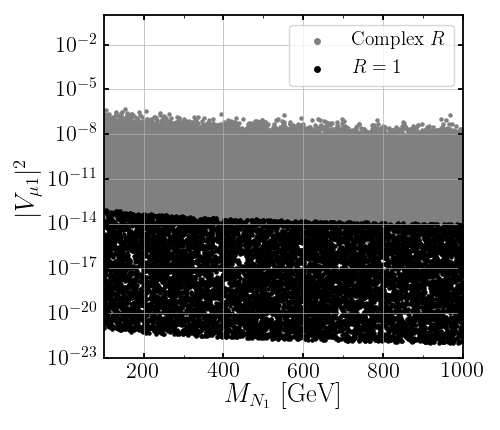}
\includegraphics[width=0.325\linewidth]{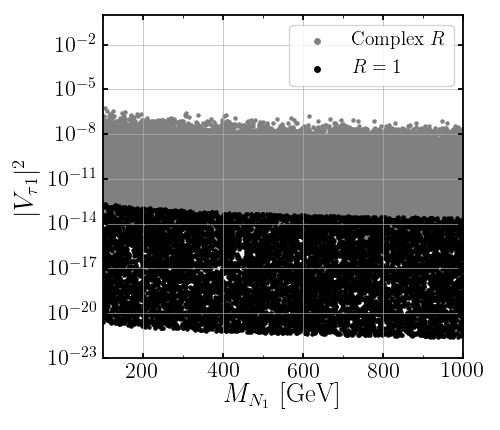}
\includegraphics[width=0.325\linewidth]{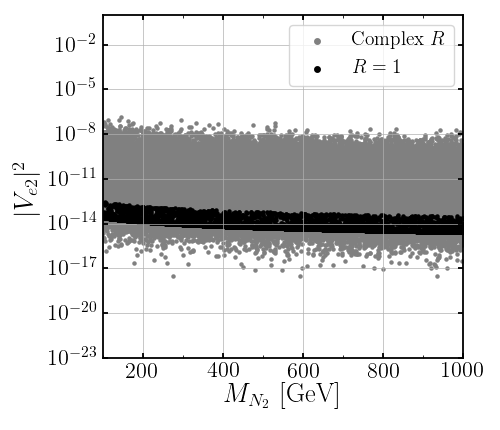}
\includegraphics[width=0.325\linewidth]{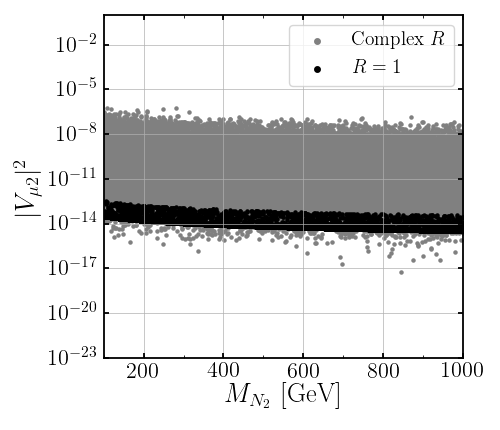}
\includegraphics[width=0.325\linewidth]{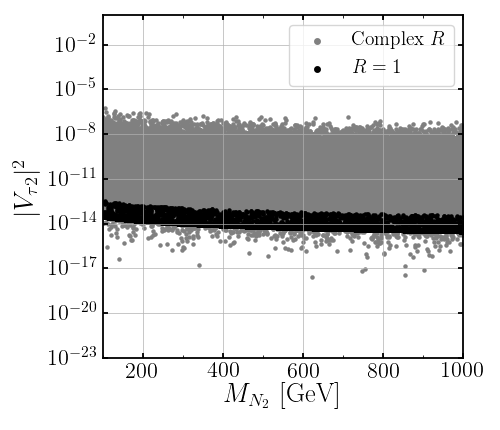}
\includegraphics[width=0.325\linewidth]{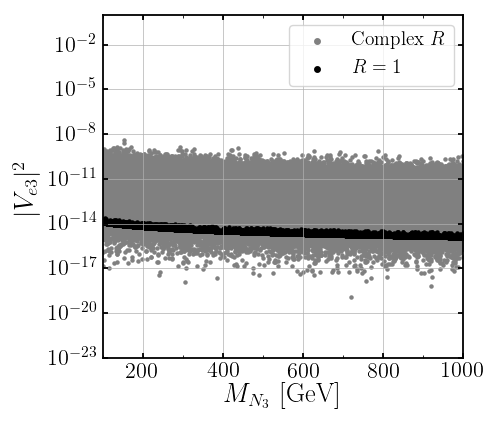}
\includegraphics[width=0.325\linewidth]{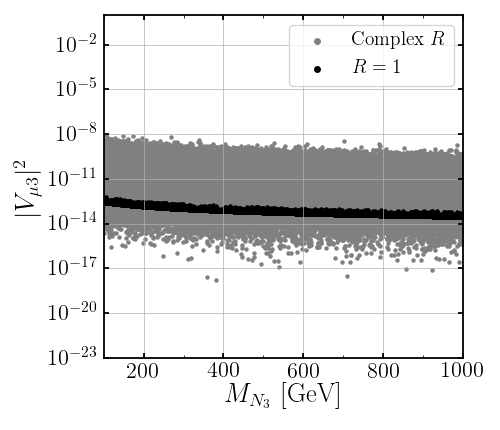}
\includegraphics[width=0.325\linewidth]{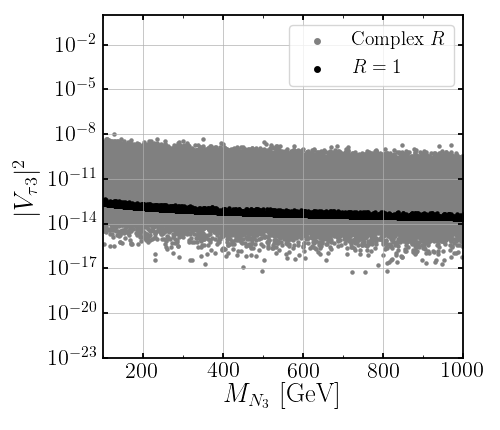}
\caption{Scatter plot of the elements of the active-sterile neutrino mixing. The black points correspond to the case with $R=1$, while the gray points correspond to a random scan on the entries of the $R$ matrix. These plots correspond to the case with normal hierarchy, for the inverted hierarchy the plots have a similar behavior with the exchange $N_1\leftrightarrow N_3$.}
\label{fig:mixings}
\end{figure}

In Fig.~\ref{fig:mixings} we present the results for the neutrino mixing matrix $V$; the black dots correspond to the simple scenario with $R=1$ and the gray points correspond to the scan over the free parameters in the ranges shown in Table~\ref{tab:ranges}.

\bibliography{DiracMajorana}

\begin{thebibliography}{41}%
\makeatletter
\providecommand \@ifxundefined [1]{%
 \@ifx{#1\undefined}
}%
\providecommand \@ifnum [1]{%
 \ifnum #1\expandafter \@firstoftwo
 \else \expandafter \@secondoftwo
 \fi
}%
\providecommand \@ifx [1]{%
 \ifx #1\expandafter \@firstoftwo
 \else \expandafter \@secondoftwo
 \fi
}%
\providecommand \natexlab [1]{#1}%
\providecommand \enquote  [1]{``#1''}%
\providecommand \bibnamefont  [1]{#1}%
\providecommand \bibfnamefont [1]{#1}%
\providecommand \citenamefont [1]{#1}%
\providecommand \href@noop [0]{\@secondoftwo}%
\providecommand \href [0]{\begingroup \@sanitize@url \@href}%
\providecommand \@href[1]{\@@startlink{#1}\@@href}%
\providecommand \@@href[1]{\endgroup#1\@@endlink}%
\providecommand \@sanitize@url [0]{\catcode `\\12\catcode `\$12\catcode
  `\&12\catcode `\#12\catcode `\^12\catcode `\_12\catcode `\%12\relax}%
\providecommand \@@startlink[1]{}%
\providecommand \@@endlink[0]{}%
\providecommand \url  [0]{\begingroup\@sanitize@url \@url }%
\providecommand \@url [1]{\endgroup\@href {#1}{\urlprefix }}%
\providecommand \urlprefix  [0]{URL }%
\providecommand \Eprint [0]{\href }%
\providecommand \doibase [0]{http://dx.doi.org/}%
\providecommand \selectlanguage [0]{\@gobble}%
\providecommand \bibinfo  [0]{\@secondoftwo}%
\providecommand \bibfield  [0]{\@secondoftwo}%
\providecommand \translation [1]{[#1]}%
\providecommand \BibitemOpen [0]{}%
\providecommand \bibitemStop [0]{}%
\providecommand \bibitemNoStop [0]{.\EOS\space}%
\providecommand \EOS [0]{\spacefactor3000\relax}%
\providecommand \BibitemShut  [1]{\csname bibitem#1\endcsname}%
\let\auto@bib@innerbib\@empty
\bibitem [{\citenamefont {Tanabashi}\ \emph {et~al.}(2018)\citenamefont
  {Tanabashi} \emph {et~al.}}]{Tanabashi:2018oca}%
  \BibitemOpen
  \bibfield  {author} {\bibinfo {author} {\bibfnamefont {M.}~\bibnamefont
  {Tanabashi}} \emph {et~al.} (\bibinfo {collaboration} {Particle Data
  Group}),\ }\bibfield  {title} {\enquote {\bibinfo {title} {{Review of
  Particle Physics}},}\ }\href {\doibase 10.1103/PhysRevD.98.030001} {\bibfield
   {journal} {\bibinfo  {journal} {Phys. Rev. D}\ }\textbf {\bibinfo {volume}
  {98}},\ \bibinfo {pages} {030001} (\bibinfo {year} {2018})}\BibitemShut
  {NoStop}%
\bibitem [{\citenamefont {Majorana}(1937)}]{Majorana:1937vz}%
  \BibitemOpen
  \bibfield  {author} {\bibinfo {author} {\bibfnamefont {E.}~\bibnamefont
  {Majorana}},\ }\bibfield  {title} {\enquote {\bibinfo {title} {{Teoria
  simmetrica dell'elettrone e del positrone}},}\ }\href {\doibase
  10.1007/BF02961314} {\bibfield  {journal} {\bibinfo  {journal} {Nuovo Cim.}\
  }\textbf {\bibinfo {volume} {14}},\ \bibinfo {pages} {171--184} (\bibinfo
  {year} {1937})}\BibitemShut {NoStop}%
\bibitem [{\citenamefont {Racah}(1937)}]{Racah:1937qq}%
  \BibitemOpen
  \bibfield  {author} {\bibinfo {author} {\bibfnamefont {G.}~\bibnamefont
  {Racah}},\ }\bibfield  {title} {\enquote {\bibinfo {title} {{On the symmetry
  of particle and antiparticle}},}\ }\href {\doibase 10.1007/BF02961321}
  {\bibfield  {journal} {\bibinfo  {journal} {Nuovo Cim.}\ }\textbf {\bibinfo
  {volume} {14}},\ \bibinfo {pages} {322--328} (\bibinfo {year}
  {1937})}\BibitemShut {NoStop}%
\bibitem [{\citenamefont {Furry}(1939)}]{Furry:1939qr}%
  \BibitemOpen
  \bibfield  {author} {\bibinfo {author} {\bibfnamefont {W.}~\bibnamefont
  {Furry}},\ }\bibfield  {title} {\enquote {\bibinfo {title} {{On transition
  probabilities in double beta-disintegration}},}\ }\href {\doibase
  10.1103/PhysRev.56.1184} {\bibfield  {journal} {\bibinfo  {journal} {Phys.
  Rev.}\ }\textbf {\bibinfo {volume} {56}},\ \bibinfo {pages} {1184--1193}
  (\bibinfo {year} {1939})}\BibitemShut {NoStop}%
\bibitem [{\citenamefont {Keung}\ and\ \citenamefont
  {Senjanovic}(1983)}]{Keung:1983uu}%
  \BibitemOpen
  \bibfield  {author} {\bibinfo {author} {\bibfnamefont {W.-Y.}\ \bibnamefont
  {Keung}}\ and\ \bibinfo {author} {\bibfnamefont {G.}~\bibnamefont
  {Senjanovic}},\ }\bibfield  {title} {\enquote {\bibinfo {title} {{Majorana
  Neutrinos and the Production of the Right-handed Charged Gauge Boson}},}\
  }\href {\doibase 10.1103/PhysRevLett.50.1427} {\bibfield  {journal} {\bibinfo
   {journal} {Phys. Rev. Lett.}\ }\textbf {\bibinfo {volume} {50}},\ \bibinfo
  {pages} {1427} (\bibinfo {year} {1983})}\BibitemShut {NoStop}%
\bibitem [{\citenamefont {Minkowski}(1977)}]{Minkowski:1977sc}%
  \BibitemOpen
  \bibfield  {author} {\bibinfo {author} {\bibfnamefont {P.}~\bibnamefont
  {Minkowski}},\ }\bibfield  {title} {\enquote {\bibinfo {title} {{$\mu \to
  e\gamma$ at a Rate of One Out of $10^{9}$ Muon Decays?}}}\ }\href {\doibase
  10.1016/0370-2693(77)90435-X} {\bibfield  {journal} {\bibinfo  {journal}
  {Phys. Lett. B}\ }\textbf {\bibinfo {volume} {67}},\ \bibinfo {pages}
  {421--428} (\bibinfo {year} {1977})}\BibitemShut {NoStop}%
\bibitem [{\citenamefont {Gell-Mann}\ \emph {et~al.}(1979)\citenamefont
  {Gell-Mann}, \citenamefont {Ramond},\ and\ \citenamefont
  {Slansky}}]{GellMann:1980vs}%
  \BibitemOpen
  \bibfield  {author} {\bibinfo {author} {\bibfnamefont {M.}~\bibnamefont
  {Gell-Mann}}, \bibinfo {author} {\bibfnamefont {P.}~\bibnamefont {Ramond}}, \
  and\ \bibinfo {author} {\bibfnamefont {R.}~\bibnamefont {Slansky}},\
  }\bibfield  {title} {\enquote {\bibinfo {title} {{Complex Spinors and Unified
  Theories}},}\ }\href@noop {} {\bibfield  {journal} {\bibinfo  {journal}
  {Conf. Proc. C}\ }\textbf {\bibinfo {volume} {790927}},\ \bibinfo {pages}
  {315--321} (\bibinfo {year} {1979})},\ \Eprint
  {http://arxiv.org/abs/1306.4669} {arXiv:1306.4669 [hep-th]} \BibitemShut
  {NoStop}%
\bibitem [{\citenamefont {Mohapatra}\ and\ \citenamefont
  {Senjanovic}(1980)}]{Mohapatra:1979ia}%
  \BibitemOpen
  \bibfield  {author} {\bibinfo {author} {\bibfnamefont {R.~N.}\ \bibnamefont
  {Mohapatra}}\ and\ \bibinfo {author} {\bibfnamefont {G.}~\bibnamefont
  {Senjanovic}},\ }\bibfield  {title} {\enquote {\bibinfo {title} {{Neutrino
  Mass and Spontaneous Parity Nonconservation}},}\ }\href {\doibase
  10.1103/PhysRevLett.44.912} {\bibfield  {journal} {\bibinfo  {journal} {Phys.
  Rev. Lett.}\ }\textbf {\bibinfo {volume} {44}},\ \bibinfo {pages} {912}
  (\bibinfo {year} {1980})}\BibitemShut {NoStop}%
\bibitem [{\citenamefont {Yanagida}(1979)}]{Yanagida:1979as}%
  \BibitemOpen
  \bibfield  {author} {\bibinfo {author} {\bibfnamefont {T.}~\bibnamefont
  {Yanagida}},\ }\bibfield  {title} {\enquote {\bibinfo {title} {{Horizontal
  gauge symmetry and masses of neutrinos}},}\ }\href@noop {} {\bibfield
  {journal} {\bibinfo  {journal} {Conf. Proc. C}\ }\textbf {\bibinfo {volume}
  {7902131}},\ \bibinfo {pages} {95--99} (\bibinfo {year} {1979})}\BibitemShut
  {NoStop}%
\bibitem [{\citenamefont {Han}\ and\ \citenamefont {Zhang}(2006)}]{Han:2006ip}%
  \BibitemOpen
  \bibfield  {author} {\bibinfo {author} {\bibfnamefont {T.}~\bibnamefont
  {Han}}\ and\ \bibinfo {author} {\bibfnamefont {B.}~\bibnamefont {Zhang}},\
  }\bibfield  {title} {\enquote {\bibinfo {title} {{Signatures for Majorana
  neutrinos at hadron colliders}},}\ }\href {\doibase
  10.1103/PhysRevLett.97.171804} {\bibfield  {journal} {\bibinfo  {journal}
  {Phys. Rev. Lett.}\ }\textbf {\bibinfo {volume} {97}},\ \bibinfo {pages}
  {171804} (\bibinfo {year} {2006})},\ \Eprint
  {http://arxiv.org/abs/hep-ph/0604064} {arXiv:hep-ph/0604064} \BibitemShut
  {NoStop}%
\bibitem [{\citenamefont {Kersten}\ and\ \citenamefont
  {Smirnov}(2007)}]{Kersten:2007vk}%
  \BibitemOpen
  \bibfield  {author} {\bibinfo {author} {\bibfnamefont {J.}~\bibnamefont
  {Kersten}}\ and\ \bibinfo {author} {\bibfnamefont {A.~Y.}\ \bibnamefont
  {Smirnov}},\ }\bibfield  {title} {\enquote {\bibinfo {title} {{Right-Handed
  Neutrinos at CERN LHC and the Mechanism of Neutrino Mass Generation}},}\
  }\href {\doibase 10.1103/PhysRevD.76.073005} {\bibfield  {journal} {\bibinfo
  {journal} {Phys. Rev. D}\ }\textbf {\bibinfo {volume} {76}},\ \bibinfo
  {pages} {073005} (\bibinfo {year} {2007})},\ \Eprint
  {http://arxiv.org/abs/0705.3221} {arXiv:0705.3221 [hep-ph]} \BibitemShut
  {NoStop}%
\bibitem [{\citenamefont {del Aguila}\ and\ \citenamefont
  {Aguilar-Saavedra}(2009{\natexlab{a}})}]{delAguila:2008cj}%
  \BibitemOpen
  \bibfield  {author} {\bibinfo {author} {\bibfnamefont {F.}~\bibnamefont {del
  Aguila}}\ and\ \bibinfo {author} {\bibfnamefont {J.}~\bibnamefont
  {Aguilar-Saavedra}},\ }\bibfield  {title} {\enquote {\bibinfo {title}
  {{Distinguishing seesaw models at LHC with multi-lepton signals}},}\ }\href
  {\doibase 10.1016/j.nuclphysb.2008.12.029} {\bibfield  {journal} {\bibinfo
  {journal} {Nucl. Phys. B}\ }\textbf {\bibinfo {volume} {813}},\ \bibinfo
  {pages} {22--90} (\bibinfo {year} {2009}{\natexlab{a}})},\ \Eprint
  {http://arxiv.org/abs/0808.2468} {arXiv:0808.2468 [hep-ph]} \BibitemShut
  {NoStop}%
\bibitem [{\citenamefont {del Aguila}\ and\ \citenamefont
  {Aguilar-Saavedra}(2009{\natexlab{b}})}]{delAguila:2008hw}%
  \BibitemOpen
  \bibfield  {author} {\bibinfo {author} {\bibfnamefont {F.}~\bibnamefont {del
  Aguila}}\ and\ \bibinfo {author} {\bibfnamefont {J.}~\bibnamefont
  {Aguilar-Saavedra}},\ }\bibfield  {title} {\enquote {\bibinfo {title}
  {{Electroweak scale seesaw and heavy Dirac neutrino signals at LHC}},}\
  }\href {\doibase 10.1016/j.physletb.2009.01.010} {\bibfield  {journal}
  {\bibinfo  {journal} {Phys. Lett. B}\ }\textbf {\bibinfo {volume} {672}},\
  \bibinfo {pages} {158--165} (\bibinfo {year} {2009}{\natexlab{b}})},\ \Eprint
  {http://arxiv.org/abs/0809.2096} {arXiv:0809.2096 [hep-ph]} \BibitemShut
  {NoStop}%
\bibitem [{\citenamefont {Atre}\ \emph {et~al.}(2009)\citenamefont {Atre},
  \citenamefont {Han}, \citenamefont {Pascoli},\ and\ \citenamefont
  {Zhang}}]{Atre:2009rg}%
  \BibitemOpen
  \bibfield  {author} {\bibinfo {author} {\bibfnamefont {A.}~\bibnamefont
  {Atre}}, \bibinfo {author} {\bibfnamefont {T.}~\bibnamefont {Han}}, \bibinfo
  {author} {\bibfnamefont {S.}~\bibnamefont {Pascoli}}, \ and\ \bibinfo
  {author} {\bibfnamefont {B.}~\bibnamefont {Zhang}},\ }\bibfield  {title}
  {\enquote {\bibinfo {title} {{The Search for Heavy Majorana Neutrinos}},}\
  }\href {\doibase 10.1088/1126-6708/2009/05/030} {\bibfield  {journal}
  {\bibinfo  {journal} {JHEP}\ }\textbf {\bibinfo {volume} {05}},\ \bibinfo
  {pages} {030} (\bibinfo {year} {2009})},\ \Eprint
  {http://arxiv.org/abs/0901.3589} {arXiv:0901.3589 [hep-ph]} \BibitemShut
  {NoStop}%
\bibitem [{\citenamefont {Khachatryan}\ \emph {et~al.}(2015)\citenamefont
  {Khachatryan} \emph {et~al.}}]{Khachatryan:2015gha}%
  \BibitemOpen
  \bibfield  {author} {\bibinfo {author} {\bibfnamefont {V.}~\bibnamefont
  {Khachatryan}} \emph {et~al.} (\bibinfo {collaboration} {CMS}),\ }\bibfield
  {title} {\enquote {\bibinfo {title} {{Search for heavy Majorana neutrinos in
  $\mu^\pm \mu^\pm+$ jets events in proton-proton collisions at $\sqrt{s}$ = 8
  TeV}},}\ }\href {\doibase 10.1016/j.physletb.2015.06.070} {\bibfield
  {journal} {\bibinfo  {journal} {Phys. Lett. B}\ }\textbf {\bibinfo {volume}
  {748}},\ \bibinfo {pages} {144--166} (\bibinfo {year} {2015})},\ \Eprint
  {http://arxiv.org/abs/1501.05566} {arXiv:1501.05566 [hep-ex]} \BibitemShut
  {NoStop}%
\bibitem [{\citenamefont {Aad}\ \emph {et~al.}(2015)\citenamefont {Aad} \emph
  {et~al.}}]{Aad:2015xaa}%
  \BibitemOpen
  \bibfield  {author} {\bibinfo {author} {\bibfnamefont {G.}~\bibnamefont
  {Aad}} \emph {et~al.} (\bibinfo {collaboration} {ATLAS}),\ }\bibfield
  {title} {\enquote {\bibinfo {title} {{Search for heavy Majorana neutrinos
  with the ATLAS detector in pp collisions at $ \sqrt{s}=8 $ TeV}},}\ }\href
  {\doibase 10.1007/JHEP07(2015)162} {\bibfield  {journal} {\bibinfo  {journal}
  {JHEP}\ }\textbf {\bibinfo {volume} {07}},\ \bibinfo {pages} {162} (\bibinfo
  {year} {2015})},\ \Eprint {http://arxiv.org/abs/1506.06020} {arXiv:1506.06020
  [hep-ex]} \BibitemShut {NoStop}%
\bibitem [{\citenamefont {Khachatryan}\ \emph {et~al.}(2016)\citenamefont
  {Khachatryan} \emph {et~al.}}]{Khachatryan:2016olu}%
  \BibitemOpen
  \bibfield  {author} {\bibinfo {author} {\bibfnamefont {V.}~\bibnamefont
  {Khachatryan}} \emph {et~al.} (\bibinfo {collaboration} {CMS}),\ }\bibfield
  {title} {\enquote {\bibinfo {title} {{Search for heavy Majorana neutrinos in
  e$^{±}$e$^{±}$+ jets and e$^{±}$ $\mu^{±}$+ jets events in proton-proton
  collisions at $ \sqrt{s}=8 $ TeV}},}\ }\href {\doibase
  10.1007/JHEP04(2016)169} {\bibfield  {journal} {\bibinfo  {journal} {JHEP}\
  }\textbf {\bibinfo {volume} {04}},\ \bibinfo {pages} {169} (\bibinfo {year}
  {2016})},\ \Eprint {http://arxiv.org/abs/1603.02248} {arXiv:1603.02248
  [hep-ex]} \BibitemShut {NoStop}%
\bibitem [{\citenamefont {Helo}\ \emph {et~al.}(2014)\citenamefont {Helo},
  \citenamefont {Hirsch},\ and\ \citenamefont {Kovalenko}}]{Helo:2013esa}%
  \BibitemOpen
  \bibfield  {author} {\bibinfo {author} {\bibfnamefont {J.~C.}\ \bibnamefont
  {Helo}}, \bibinfo {author} {\bibfnamefont {M.}~\bibnamefont {Hirsch}}, \ and\
  \bibinfo {author} {\bibfnamefont {S.}~\bibnamefont {Kovalenko}},\ }\bibfield
  {title} {\enquote {\bibinfo {title} {{Heavy neutrino searches at the LHC with
  displaced vertices}},}\ }\href {\doibase 10.1103/PhysRevD.89.073005}
  {\bibfield  {journal} {\bibinfo  {journal} {Phys. Rev. D}\ }\textbf {\bibinfo
  {volume} {89}},\ \bibinfo {pages} {073005} (\bibinfo {year} {2014})},\
  \bibinfo {note} {[Erratum: Phys.Rev.D 93, 099902 (2016)]},\ \Eprint
  {http://arxiv.org/abs/1312.2900} {arXiv:1312.2900 [hep-ph]} \BibitemShut
  {NoStop}%
\bibitem [{\citenamefont {Izaguirre}\ and\ \citenamefont
  {Shuve}(2015)}]{Izaguirre:2015pga}%
  \BibitemOpen
  \bibfield  {author} {\bibinfo {author} {\bibfnamefont {E.}~\bibnamefont
  {Izaguirre}}\ and\ \bibinfo {author} {\bibfnamefont {B.}~\bibnamefont
  {Shuve}},\ }\bibfield  {title} {\enquote {\bibinfo {title} {{Multilepton and
  Lepton Jet Probes of Sub-Weak-Scale Right-Handed Neutrinos}},}\ }\href
  {\doibase 10.1103/PhysRevD.91.093010} {\bibfield  {journal} {\bibinfo
  {journal} {Phys. Rev. D}\ }\textbf {\bibinfo {volume} {91}},\ \bibinfo
  {pages} {093010} (\bibinfo {year} {2015})},\ \Eprint
  {http://arxiv.org/abs/1504.02470} {arXiv:1504.02470 [hep-ph]} \BibitemShut
  {NoStop}%
\bibitem [{\citenamefont {Batell}\ \emph {et~al.}(2016)\citenamefont {Batell},
  \citenamefont {Pospelov},\ and\ \citenamefont {Shuve}}]{Batell:2016zod}%
  \BibitemOpen
  \bibfield  {author} {\bibinfo {author} {\bibfnamefont {B.}~\bibnamefont
  {Batell}}, \bibinfo {author} {\bibfnamefont {M.}~\bibnamefont {Pospelov}}, \
  and\ \bibinfo {author} {\bibfnamefont {B.}~\bibnamefont {Shuve}},\ }\bibfield
   {title} {\enquote {\bibinfo {title} {{Shedding Light on Neutrino Masses with
  Dark Forces}},}\ }\href {\doibase 10.1007/JHEP08(2016)052} {\bibfield
  {journal} {\bibinfo  {journal} {JHEP}\ }\textbf {\bibinfo {volume} {08}},\
  \bibinfo {pages} {052} (\bibinfo {year} {2016})},\ \Eprint
  {http://arxiv.org/abs/1604.06099} {arXiv:1604.06099 [hep-ph]} \BibitemShut
  {NoStop}%
\bibitem [{\citenamefont {Graesser}(2007)}]{Graesser:2007yj}%
  \BibitemOpen
  \bibfield  {author} {\bibinfo {author} {\bibfnamefont {M.~L.}\ \bibnamefont
  {Graesser}},\ }\bibfield  {title} {\enquote {\bibinfo {title} {{Broadening
  the Higgs boson with right-handed neutrinos and a higher dimension operator
  at the electroweak scale}},}\ }\href {\doibase 10.1103/PhysRevD.76.075006}
  {\bibfield  {journal} {\bibinfo  {journal} {Phys. Rev. D}\ }\textbf {\bibinfo
  {volume} {76}},\ \bibinfo {pages} {075006} (\bibinfo {year} {2007})},\
  \Eprint {http://arxiv.org/abs/0704.0438} {arXiv:0704.0438 [hep-ph]}
  \BibitemShut {NoStop}%
\bibitem [{\citenamefont {Caputo}\ \emph {et~al.}(2017)\citenamefont {Caputo},
  \citenamefont {Hernandez}, \citenamefont {Lopez-Pavon},\ and\ \citenamefont
  {Salvado}}]{Caputo:2017pit}%
  \BibitemOpen
  \bibfield  {author} {\bibinfo {author} {\bibfnamefont {A.}~\bibnamefont
  {Caputo}}, \bibinfo {author} {\bibfnamefont {P.}~\bibnamefont {Hernandez}},
  \bibinfo {author} {\bibfnamefont {J.}~\bibnamefont {Lopez-Pavon}}, \ and\
  \bibinfo {author} {\bibfnamefont {J.}~\bibnamefont {Salvado}},\ }\bibfield
  {title} {\enquote {\bibinfo {title} {{The seesaw portal in testable models of
  neutrino masses}},}\ }\href {\doibase 10.1007/JHEP06(2017)112} {\bibfield
  {journal} {\bibinfo  {journal} {JHEP}\ }\textbf {\bibinfo {volume} {06}},\
  \bibinfo {pages} {112} (\bibinfo {year} {2017})},\ \Eprint
  {http://arxiv.org/abs/1704.08721} {arXiv:1704.08721 [hep-ph]} \BibitemShut
  {NoStop}%
\bibitem [{\citenamefont {Deppisch}\ \emph {et~al.}(2018)\citenamefont
  {Deppisch}, \citenamefont {Liu},\ and\ \citenamefont
  {Mitra}}]{Deppisch:2018eth}%
  \BibitemOpen
  \bibfield  {author} {\bibinfo {author} {\bibfnamefont {F.~F.}\ \bibnamefont
  {Deppisch}}, \bibinfo {author} {\bibfnamefont {W.}~\bibnamefont {Liu}}, \
  and\ \bibinfo {author} {\bibfnamefont {M.}~\bibnamefont {Mitra}},\ }\bibfield
   {title} {\enquote {\bibinfo {title} {{Long-lived Heavy Neutrinos from Higgs
  Decays}},}\ }\href {\doibase 10.1007/JHEP08(2018)181} {\bibfield  {journal}
  {\bibinfo  {journal} {JHEP}\ }\textbf {\bibinfo {volume} {08}},\ \bibinfo
  {pages} {181} (\bibinfo {year} {2018})},\ \Eprint
  {http://arxiv.org/abs/1804.04075} {arXiv:1804.04075 [hep-ph]} \BibitemShut
  {NoStop}%
\bibitem [{\citenamefont {Butterworth}\ \emph {et~al.}(2019)\citenamefont
  {Butterworth}, \citenamefont {Chala}, \citenamefont {Englert}, \citenamefont
  {Spannowsky},\ and\ \citenamefont {Titov}}]{Butterworth:2019iff}%
  \BibitemOpen
  \bibfield  {author} {\bibinfo {author} {\bibfnamefont {J.~M.}\ \bibnamefont
  {Butterworth}}, \bibinfo {author} {\bibfnamefont {M.}~\bibnamefont {Chala}},
  \bibinfo {author} {\bibfnamefont {C.}~\bibnamefont {Englert}}, \bibinfo
  {author} {\bibfnamefont {M.}~\bibnamefont {Spannowsky}}, \ and\ \bibinfo
  {author} {\bibfnamefont {A.}~\bibnamefont {Titov}},\ }\bibfield  {title}
  {\enquote {\bibinfo {title} {{Higgs phenomenology as a probe of sterile
  neutrinos}},}\ }\href {\doibase 10.1103/PhysRevD.100.115019} {\bibfield
  {journal} {\bibinfo  {journal} {Phys. Rev. D}\ }\textbf {\bibinfo {volume}
  {100}},\ \bibinfo {pages} {115019} (\bibinfo {year} {2019})},\ \Eprint
  {http://arxiv.org/abs/1909.04665} {arXiv:1909.04665 [hep-ph]} \BibitemShut
  {NoStop}%
\bibitem [{\citenamefont {Cai}\ \emph {et~al.}(2018)\citenamefont {Cai},
  \citenamefont {Han}, \citenamefont {Li},\ and\ \citenamefont
  {Ruiz}}]{Cai:2017mow}%
  \BibitemOpen
  \bibfield  {author} {\bibinfo {author} {\bibfnamefont {Y.}~\bibnamefont
  {Cai}}, \bibinfo {author} {\bibfnamefont {T.}~\bibnamefont {Han}}, \bibinfo
  {author} {\bibfnamefont {T.}~\bibnamefont {Li}}, \ and\ \bibinfo {author}
  {\bibfnamefont {R.}~\bibnamefont {Ruiz}},\ }\bibfield  {title} {\enquote
  {\bibinfo {title} {{Lepton Number Violation: Seesaw Models and Their Collider
  Tests}},}\ }\href {\doibase 10.3389/fphy.2018.00040} {\bibfield  {journal}
  {\bibinfo  {journal} {Front. in Phys.}\ }\textbf {\bibinfo {volume} {6}},\
  \bibinfo {pages} {40} (\bibinfo {year} {2018})},\ \Eprint
  {http://arxiv.org/abs/1711.02180} {arXiv:1711.02180 [hep-ph]} \BibitemShut
  {NoStop}%
\bibitem [{\citenamefont {Feldman}\ \emph {et~al.}(2012)\citenamefont
  {Feldman}, \citenamefont {Fileviez~Perez},\ and\ \citenamefont
  {Nath}}]{Feldman:2011ms}%
  \BibitemOpen
  \bibfield  {author} {\bibinfo {author} {\bibfnamefont {D.}~\bibnamefont
  {Feldman}}, \bibinfo {author} {\bibfnamefont {P.}~\bibnamefont
  {Fileviez~Perez}}, \ and\ \bibinfo {author} {\bibfnamefont {P.}~\bibnamefont
  {Nath}},\ }\bibfield  {title} {\enquote {\bibinfo {title} {{R-parity
  Conservation via the Stueckelberg Mechanism: LHC and Dark Matter Signals}},}\
  }\href {\doibase 10.1007/JHEP01(2012)038} {\bibfield  {journal} {\bibinfo
  {journal} {JHEP}\ }\textbf {\bibinfo {volume} {01}},\ \bibinfo {pages} {038}
  (\bibinfo {year} {2012})},\ \Eprint {http://arxiv.org/abs/1109.2901}
  {arXiv:1109.2901 [hep-ph]} \BibitemShut {NoStop}%
\bibitem [{\citenamefont {Fileviez~Perez}\ \emph {et~al.}(2009)\citenamefont
  {Fileviez~Perez}, \citenamefont {Han},\ and\ \citenamefont
  {Li}}]{Perez:2009mu}%
  \BibitemOpen
  \bibfield  {author} {\bibinfo {author} {\bibfnamefont {P.}~\bibnamefont
  {Fileviez~Perez}}, \bibinfo {author} {\bibfnamefont {T.}~\bibnamefont {Han}},
  \ and\ \bibinfo {author} {\bibfnamefont {T.}~\bibnamefont {Li}},\ }\bibfield
  {title} {\enquote {\bibinfo {title} {{Testability of Type I Seesaw at the
  CERN LHC: Revealing the Existence of the B-L Symmetry}},}\ }\href {\doibase
  10.1103/PhysRevD.80.073015} {\bibfield  {journal} {\bibinfo  {journal} {Phys.
  Rev. D}\ }\textbf {\bibinfo {volume} {80}},\ \bibinfo {pages} {073015}
  (\bibinfo {year} {2009})},\ \Eprint {http://arxiv.org/abs/0907.4186}
  {arXiv:0907.4186 [hep-ph]} \BibitemShut {NoStop}%
\bibitem [{\citenamefont {Huitu}\ \emph {et~al.}(2008)\citenamefont {Huitu},
  \citenamefont {Khalil}, \citenamefont {Okada},\ and\ \citenamefont
  {Rai}}]{Huitu:2008gf}%
  \BibitemOpen
  \bibfield  {author} {\bibinfo {author} {\bibfnamefont {K.}~\bibnamefont
  {Huitu}}, \bibinfo {author} {\bibfnamefont {S.}~\bibnamefont {Khalil}},
  \bibinfo {author} {\bibfnamefont {H.}~\bibnamefont {Okada}}, \ and\ \bibinfo
  {author} {\bibfnamefont {S.~K.}\ \bibnamefont {Rai}},\ }\bibfield  {title}
  {\enquote {\bibinfo {title} {{Signatures for right-handed neutrinos at the
  Large Hadron Collider}},}\ }\href {\doibase 10.1103/PhysRevLett.101.181802}
  {\bibfield  {journal} {\bibinfo  {journal} {Phys. Rev. Lett.}\ }\textbf
  {\bibinfo {volume} {101}},\ \bibinfo {pages} {181802} (\bibinfo {year}
  {2008})},\ \Eprint {http://arxiv.org/abs/0803.2799} {arXiv:0803.2799
  [hep-ph]} \BibitemShut {NoStop}%
\bibitem [{\citenamefont {Basso}\ \emph {et~al.}(2009)\citenamefont {Basso},
  \citenamefont {Belyaev}, \citenamefont {Moretti},\ and\ \citenamefont
  {Shepherd-Themistocleous}}]{Basso:2008iv}%
  \BibitemOpen
  \bibfield  {author} {\bibinfo {author} {\bibfnamefont {L.}~\bibnamefont
  {Basso}}, \bibinfo {author} {\bibfnamefont {A.}~\bibnamefont {Belyaev}},
  \bibinfo {author} {\bibfnamefont {S.}~\bibnamefont {Moretti}}, \ and\
  \bibinfo {author} {\bibfnamefont {C.~H.}\ \bibnamefont
  {Shepherd-Themistocleous}},\ }\bibfield  {title} {\enquote {\bibinfo {title}
  {{Phenomenology of the minimal B-L extension of the Standard model: Z' and
  neutrinos}},}\ }\href {\doibase 10.1103/PhysRevD.80.055030} {\bibfield
  {journal} {\bibinfo  {journal} {Phys. Rev. D}\ }\textbf {\bibinfo {volume}
  {80}},\ \bibinfo {pages} {055030} (\bibinfo {year} {2009})},\ \Eprint
  {http://arxiv.org/abs/0812.4313} {arXiv:0812.4313 [hep-ph]} \BibitemShut
  {NoStop}%
\bibitem [{\citenamefont {Kang}\ \emph {et~al.}(2016)\citenamefont {Kang},
  \citenamefont {Ko},\ and\ \citenamefont {Li}}]{Kang:2015uoc}%
  \BibitemOpen
  \bibfield  {author} {\bibinfo {author} {\bibfnamefont {Z.}~\bibnamefont
  {Kang}}, \bibinfo {author} {\bibfnamefont {P.}~\bibnamefont {Ko}}, \ and\
  \bibinfo {author} {\bibfnamefont {J.}~\bibnamefont {Li}},\ }\bibfield
  {title} {\enquote {\bibinfo {title} {{New Avenues to Heavy Right-handed
  Neutrinos with Pair Production at Hadronic Colliders}},}\ }\href {\doibase
  10.1103/PhysRevD.93.075037} {\bibfield  {journal} {\bibinfo  {journal} {Phys.
  Rev. D}\ }\textbf {\bibinfo {volume} {93}},\ \bibinfo {pages} {075037}
  (\bibinfo {year} {2016})},\ \Eprint {http://arxiv.org/abs/1512.08373}
  {arXiv:1512.08373 [hep-ph]} \BibitemShut {NoStop}%
\bibitem [{\citenamefont {Alioli}\ \emph {et~al.}(2018)\citenamefont {Alioli},
  \citenamefont {Farina}, \citenamefont {Pappadopulo},\ and\ \citenamefont
  {Ruderman}}]{Alioli:2017nzr}%
  \BibitemOpen
  \bibfield  {author} {\bibinfo {author} {\bibfnamefont {S.}~\bibnamefont
  {Alioli}}, \bibinfo {author} {\bibfnamefont {M.}~\bibnamefont {Farina}},
  \bibinfo {author} {\bibfnamefont {D.}~\bibnamefont {Pappadopulo}}, \ and\
  \bibinfo {author} {\bibfnamefont {J.~T.}\ \bibnamefont {Ruderman}},\
  }\bibfield  {title} {\enquote {\bibinfo {title} {{Catching a New Force by the
  Tail}},}\ }\href {\doibase 10.1103/PhysRevLett.120.101801} {\bibfield
  {journal} {\bibinfo  {journal} {Phys.\ Rev.\ Lett.}\ }\textbf {\bibinfo
  {volume} {120}},\ \bibinfo {pages} {101801} (\bibinfo {year} {2018})},\
  \Eprint {http://arxiv.org/abs/1712.02347} {arXiv:1712.02347 [hep-ph]}
  \BibitemShut {NoStop}%
\bibitem [{\citenamefont {Aaboud}\ \emph {et~al.}(2017)\citenamefont {Aaboud}
  \emph {et~al.}}]{Aaboud:2017buh}%
  \BibitemOpen
  \bibfield  {author} {\bibinfo {author} {\bibfnamefont {M.}~\bibnamefont
  {Aaboud}} \emph {et~al.} (\bibinfo {collaboration} {ATLAS}),\ }\bibfield
  {title} {\enquote {\bibinfo {title} {{Search for new high-mass phenomena in
  the dilepton final state using 36 fb$^{−1}$ of proton-proton collision data
  at $ \sqrt{s}=13 $ TeV with the ATLAS detector}},}\ }\href {\doibase
  10.1007/JHEP10(2017)182} {\bibfield  {journal} {\bibinfo  {journal} {JHEP}\
  }\textbf {\bibinfo {volume} {10}},\ \bibinfo {pages} {182} (\bibinfo {year}
  {2017})},\ \Eprint {http://arxiv.org/abs/1707.02424} {arXiv:1707.02424
  [hep-ex]} \BibitemShut {NoStop}%
\bibitem [{\citenamefont {Fileviez~Pérez}\ \emph {et~al.}(2019)\citenamefont
  {Fileviez~Pérez}, \citenamefont {Murgui},\ and\ \citenamefont
  {Plascencia}}]{FileviezPerez:2019cyn}%
  \BibitemOpen
  \bibfield  {author} {\bibinfo {author} {\bibfnamefont {P.}~\bibnamefont
  {Fileviez~Pérez}}, \bibinfo {author} {\bibfnamefont {C.}~\bibnamefont
  {Murgui}}, \ and\ \bibinfo {author} {\bibfnamefont {A.~D.}\ \bibnamefont
  {Plascencia}},\ }\bibfield  {title} {\enquote {\bibinfo {title}
  {{Neutrino-Dark Matter Connections in Gauge Theories}},}\ }\href {\doibase
  10.1103/PhysRevD.100.035041} {\bibfield  {journal} {\bibinfo  {journal}
  {Phys. Rev. D}\ }\textbf {\bibinfo {volume} {100}},\ \bibinfo {pages}
  {035041} (\bibinfo {year} {2019})},\ \Eprint
  {http://arxiv.org/abs/1905.06344} {arXiv:1905.06344 [hep-ph]} \BibitemShut
  {NoStop}%
\bibitem [{\citenamefont {Fileviez~Perez}\ and\ \citenamefont
  {Murgui}(2018)}]{Perez:2017qns}%
  \BibitemOpen
  \bibfield  {author} {\bibinfo {author} {\bibfnamefont {P.}~\bibnamefont
  {Fileviez~Perez}}\ and\ \bibinfo {author} {\bibfnamefont {C.}~\bibnamefont
  {Murgui}},\ }\bibfield  {title} {\enquote {\bibinfo {title} {{Sterile
  neutrinos and B--L symmetry}},}\ }\href {\doibase
  10.1016/j.physletb.2017.12.041} {\bibfield  {journal} {\bibinfo  {journal}
  {Phys.\ Lett.\ B}\ }\textbf {\bibinfo {volume} {777}},\ \bibinfo {pages}
  {381--387} (\bibinfo {year} {2018})},\ \Eprint
  {http://arxiv.org/abs/1708.02247} {arXiv:1708.02247 [hep-ph]} \BibitemShut
  {NoStop}%
\bibitem [{\citenamefont {Martin}\ \emph {et~al.}(2009)\citenamefont {Martin},
  \citenamefont {Stirling}, \citenamefont {Thorne},\ and\ \citenamefont
  {Watt}}]{Martin:2009iq}%
  \BibitemOpen
  \bibfield  {author} {\bibinfo {author} {\bibfnamefont {A.}~\bibnamefont
  {Martin}}, \bibinfo {author} {\bibfnamefont {W.}~\bibnamefont {Stirling}},
  \bibinfo {author} {\bibfnamefont {R.}~\bibnamefont {Thorne}}, \ and\ \bibinfo
  {author} {\bibfnamefont {G.}~\bibnamefont {Watt}},\ }\bibfield  {title}
  {\enquote {\bibinfo {title} {{Parton distributions for the LHC}},}\ }\href
  {\doibase 10.1140/epjc/s10052-009-1072-5} {\bibfield  {journal} {\bibinfo
  {journal} {Eur. Phys. J. C}\ }\textbf {\bibinfo {volume} {63}},\ \bibinfo
  {pages} {189--285} (\bibinfo {year} {2009})},\ \Eprint
  {http://arxiv.org/abs/0901.0002} {arXiv:0901.0002 [hep-ph]} \BibitemShut
  {NoStop}%
\bibitem [{\citenamefont {Esteban}\ \emph {et~al.}(2019)\citenamefont
  {Esteban}, \citenamefont {Gonzalez-Garcia}, \citenamefont
  {Hernandez-Cabezudo}, \citenamefont {Maltoni},\ and\ \citenamefont
  {Schwetz}}]{Esteban:2018azc}%
  \BibitemOpen
  \bibfield  {author} {\bibinfo {author} {\bibfnamefont {I.}~\bibnamefont
  {Esteban}}, \bibinfo {author} {\bibfnamefont {M.}~\bibnamefont
  {Gonzalez-Garcia}}, \bibinfo {author} {\bibfnamefont {A.}~\bibnamefont
  {Hernandez-Cabezudo}}, \bibinfo {author} {\bibfnamefont {M.}~\bibnamefont
  {Maltoni}}, \ and\ \bibinfo {author} {\bibfnamefont {T.}~\bibnamefont
  {Schwetz}},\ }\bibfield  {title} {\enquote {\bibinfo {title} {{Global
  analysis of three-flavour neutrino oscillations: synergies and tensions in
  the determination of $\theta_{23}$, $\delta_{CP}$, and the mass ordering}},}\
  }\href {\doibase 10.1007/JHEP01(2019)106} {\bibfield  {journal} {\bibinfo
  {journal} {JHEP}\ }\textbf {\bibinfo {volume} {01}},\ \bibinfo {pages} {106}
  (\bibinfo {year} {2019})},\ \Eprint {http://arxiv.org/abs/1811.05487}
  {arXiv:1811.05487 [hep-ph]} \BibitemShut {NoStop}%
\bibitem [{\citenamefont {Casas}\ and\ \citenamefont
  {Ibarra}(2001)}]{Casas:2001sr}%
  \BibitemOpen
  \bibfield  {author} {\bibinfo {author} {\bibfnamefont {J.}~\bibnamefont
  {Casas}}\ and\ \bibinfo {author} {\bibfnamefont {A.}~\bibnamefont {Ibarra}},\
  }\bibfield  {title} {\enquote {\bibinfo {title} {{Oscillating neutrinos and
  $\mu \to e \gamma$}},}\ }\href {\doibase 10.1016/S0550-3213(01)00475-8}
  {\bibfield  {journal} {\bibinfo  {journal} {Nucl. Phys. B}\ }\textbf
  {\bibinfo {volume} {618}},\ \bibinfo {pages} {171--204} (\bibinfo {year}
  {2001})},\ \Eprint {http://arxiv.org/abs/hep-ph/0103065}
  {arXiv:hep-ph/0103065} \BibitemShut {NoStop}%
\bibitem [{\citenamefont {Feng}\ \emph {et~al.}(2018)\citenamefont {Feng},
  \citenamefont {Galon}, \citenamefont {Kling},\ and\ \citenamefont
  {Trojanowski}}]{Feng:2017uoz}%
  \BibitemOpen
  \bibfield  {author} {\bibinfo {author} {\bibfnamefont {J.~L.}\ \bibnamefont
  {Feng}}, \bibinfo {author} {\bibfnamefont {I.}~\bibnamefont {Galon}},
  \bibinfo {author} {\bibfnamefont {F.}~\bibnamefont {Kling}}, \ and\ \bibinfo
  {author} {\bibfnamefont {S.}~\bibnamefont {Trojanowski}},\ }\bibfield
  {title} {\enquote {\bibinfo {title} {{ForwArd Search ExpeRiment at the
  LHC}},}\ }\href {\doibase 10.1103/PhysRevD.97.035001} {\bibfield  {journal}
  {\bibinfo  {journal} {Phys. Rev. D}\ }\textbf {\bibinfo {volume} {97}},\
  \bibinfo {pages} {035001} (\bibinfo {year} {2018})},\ \Eprint
  {http://arxiv.org/abs/1708.09389} {arXiv:1708.09389 [hep-ph]} \BibitemShut
  {NoStop}%
\bibitem [{\citenamefont {Curtin}\ \emph {et~al.}(2019)\citenamefont {Curtin}
  \emph {et~al.}}]{Curtin:2018mvb}%
  \BibitemOpen
  \bibfield  {author} {\bibinfo {author} {\bibfnamefont {D.}~\bibnamefont
  {Curtin}} \emph {et~al.},\ }\bibfield  {title} {\enquote {\bibinfo {title}
  {{Long-Lived Particles at the Energy Frontier: The MATHUSLA Physics Case}},}\
  }\href {\doibase 10.1088/1361-6633/ab28d6} {\bibfield  {journal} {\bibinfo
  {journal} {Rept. Prog. Phys.}\ }\textbf {\bibinfo {volume} {82}},\ \bibinfo
  {pages} {116201} (\bibinfo {year} {2019})},\ \Eprint
  {http://arxiv.org/abs/1806.07396} {arXiv:1806.07396 [hep-ph]} \BibitemShut
  {NoStop}%
\bibitem [{\citenamefont {Li}\ \emph {et~al.}(2009)\citenamefont {Li},
  \citenamefont {Petriello},\ and\ \citenamefont {Quackenbush}}]{Li:2009xh}%
  \BibitemOpen
  \bibfield  {author} {\bibinfo {author} {\bibfnamefont {Y.}~\bibnamefont
  {Li}}, \bibinfo {author} {\bibfnamefont {F.}~\bibnamefont {Petriello}}, \
  and\ \bibinfo {author} {\bibfnamefont {S.}~\bibnamefont {Quackenbush}},\
  }\bibfield  {title} {\enquote {\bibinfo {title} {{Reconstructing a Z-prime
  Lagrangian using the LHC and low-energy data}},}\ }\href {\doibase
  10.1103/PhysRevD.80.055018} {\bibfield  {journal} {\bibinfo  {journal} {Phys.
  Rev. D}\ }\textbf {\bibinfo {volume} {80}},\ \bibinfo {pages} {055018}
  (\bibinfo {year} {2009})},\ \Eprint {http://arxiv.org/abs/0906.4132}
  {arXiv:0906.4132 [hep-ph]} \BibitemShut {NoStop}%
\bibitem [{\citenamefont {Schael}\ \emph {et~al.}(2006)\citenamefont {Schael}
  \emph {et~al.}}]{ALEPH:2005ab}%
  \BibitemOpen
  \bibfield  {author} {\bibinfo {author} {\bibfnamefont {S.}~\bibnamefont
  {Schael}} \emph {et~al.} (\bibinfo {collaboration} {ALEPH, DELPHI, L3, OPAL,
  SLD, LEP Electroweak Working Group, SLD Electroweak Group, SLD Heavy Flavour
  Group}),\ }\bibfield  {title} {\enquote {\bibinfo {title} {{Precision
  electroweak measurements on the $Z$ resonance}},}\ }\href {\doibase
  10.1016/j.physrep.2005.12.006} {\bibfield  {journal} {\bibinfo  {journal}
  {Phys. Rept.}\ }\textbf {\bibinfo {volume} {427}},\ \bibinfo {pages}
  {257--454} (\bibinfo {year} {2006})},\ \Eprint
  {http://arxiv.org/abs/hep-ex/0509008} {arXiv:hep-ex/0509008} \BibitemShut
  {NoStop}%
\end{thebibliography}%

\end{document}